\documentclass{article}

\usepackage{arxiv}
\usepackage[utf8]{inputenc} 
\usepackage[T1]{fontenc}    
\usepackage{hyperref}       
\usepackage{url}            
\usepackage{booktabs}       
\usepackage{amsfonts}       
\usepackage[numbers]{natbib}
\usepackage{amsmath, enumitem}
\usepackage{amssymb}
\usepackage{upgreek}
\usepackage{lineno}
\usepackage{soul}
\usepackage{fixltx2e}

\usepackage{nccmath}
\usepackage{ifthen}
\usepackage{color}
\usepackage{graphicx}
\usepackage{algorithm}
\usepackage{algorithmic}
\usepackage{setspace}
\usepackage{multirow}
\graphicspath{{Fig/}}

\DeclareMathOperator*{\argmax}{argmax}

\let\oldequation\equation
\let\oldendequation\endequation
\renewenvironment{equation}
{\linenomathNonumbers\oldequation}
{\oldendequation\endlinenomath}

\title{Bayesian Inversion, Uncertainty Analysis and Interrogation using Boosting Variational Inference}
\date{} 					

\author{
	Xuebin Zhao \\
	School of Geosciences \\
	University of Edinburgh\\
	Edinburgh, Unite Kingdom \\
	\texttt{xuebin.zhao@ed.ac.uk} \\
	\And
	Andrew Curtis \\
	School of Geosciences \\
	University of Edinburgh\\
	Edinburgh, Unite Kingdom \\
	\texttt{andrew.curtis@ed.ac.uk} \\
}

\begin{document}
	\maketitle

\begin{abstract}
Geoscientists use observed data to estimate properties of the Earth's interior. This often requires non-linear inverse problems to be solved and uncertainties to be estimated. Bayesian inference solves inverse problems under a probabilistic framework, in which uncertainty is represented by a so-called posterior probability distribution. Recently, variational inference has emerged as an efficient method to estimate Bayesian solutions. By seeking the closest approximation to the posterior distribution within any chosen family of distributions, variational inference yields a fully probabilistic solution. It is important to define expressive variational families so that the posterior distribution can be represented accurately. We introduce \textit{boosting variational inference} (BVI) as a computationally efficient means to construct a flexible approximating family comprising all possible finite mixtures of simpler component distributions. We use Gaussian mixture components due to their fully parametric nature and the ease with which they can be optimised. We apply BVI to seismic travel time tomography and full waveform inversion, comparing its performance with other methods of solution. The results demonstrate that BVI achieves reasonable efficiency and accuracy while enabling the construction of a fully analytic expression for the posterior distribution. Samples that represent major components of uncertainty in the solution can be obtained analytically from each mixture component. We demonstrate that these samples can be used to solve an interrogation problem: to assess the size of a subsurface target structure. To the best of our knowledge, this is the first method in geophysics that provides both analytic and reasonably accurate probabilistic solutions to fully non-linear, high-dimensional Bayesian full waveform inversion problems. 
\end{abstract}

\section{Introduction}
In scientific investigations, the ultimate goal is usually to answer some specific, high-level questions. These can be general (\textit{what does the subsurface look like?}) but are often more specific: \textit{How large is structure X? How likely is it that this volcano will erupt within 5 days? Would another experiment be productive?} Answers to the latter questions are all one or zero-dimensional, yet in geophysical investigations of the Earth's subsurface they are often answered by interpreting high-dimensional imaging or inversion results. This often produces biased answers, first because human interpretation is a subjective process \cite{bond2007you, polson2010dynamics}, and second because uncertainties in the results are often not evaluated and seldom fully interpreted. In this work, we show that such high-level questions can be answered systematically using interrogation theory \cite{arnold2018interrogation}, which combines inverse, decision, elicitation theory and experimental design theory to optimise scientific investigations, with the overall goal to obtain the most informative answers to specific scientific inquiries.

To answer such questions accurately we need to acquire information about the Earth's interior structures. This can be obtained from data recorded either on or beneath the Earth's surface, or in the oceans, atmosphere or near-Earth orbits. Properties of interest are usually described by parameters that cannot be observed directly, and are inferred by solving an inverse (inference) problem.
In practice, inverse problem solutions are always non-unique, meaning that an infinite number of subsurface models are possible, so it is crucial to estimate the range of possible properties that are consistent with observations if solutions are to be interpreted in a reliable manner \cite{tarantola2005inverse}. Thereafter, the associated interrogation problems can be solved with minimal bias \cite{arnold2018interrogation, ely2018assessing, zhang2021interrogation, zhao2022interrogating, siahkoohi2022deep}.



A suite of methods collectively referred to as \textit{Bayesian inversion} or \textit{Bayesian inference} allow statistics of the full uncertainty structure of the inverse problem solution to be estimated. These methods employ Bayes' rule to update \textit{prior} (initial) knowledge about the parameter values that is described probabilistically, using new information provided by the observed data. The result of the inversion is represented by the \textit{posterior} probability distribution (or density) function (pdf): in principle this provides a complete solution which describes all parameter values that are consistent with the data, and quantifies their relative probabilities. 

Bayesian inference often uses global search methods such as random sampling to characterise the family of values in parameter space that yield acceptable data fits \cite{rothman1986automatic, stoffa1991nonlinear, sen2013global, sambridge1999geophysical}. Various Monte Carlo methods \cite{press1968earth, anderssen1971simple, mosegaard1995monte, gallagher2009markov, bodin2009seismic, kaufl2016solving, fichtner2018hamiltonian, izzatullah2020bayesian, khoshkholgh2021informed} have been studied extensively for different geophysical inversion problems. However, such methods still have notable issues that can become problematic in practical problems: (1) slow convergence, sometimes converging only in infinite time \cite{atchade2005adaptive, andrieu2008tutorial}, and detecting convergence of a McMC run is not straightforward \cite{gelman1992inference}; (2) poor scalability to problems with many parameters due to the curse of dimensionality \cite{scales1996uncertainties, curtis2001prior}; and (3) parallelising some methods at the sample level is not possible \cite{neiswanger2013asymptotically}.

A different approach to finding Bayesian solutions is referred to as \textit{variational inference}. In variational methods, a family of simple probability distributions is defined (often referred to as the variational family), and an optimal member of this family is sought which best approximates the true (unknown) posterior pdf. In this way, variational inference assumes a predefined (known) complexity of the posterior pdf, and hence the search space becomes more limited and clearly-defined compared to Monte Carlo methods which often assume fully unknown posterior structures. The optimal solution can be found by minimising the difference (or mathematically speaking, the distance) between the posterior and variational distributions, and the Kullback-Leibler (KL) divergence \cite{kullback1951information} is typically used for measuring this distance between two distributions. Thus, variational methods solve Bayesian problems using potentially efficient and parallelisable optimisation processes and offer well understood convergence criteria \cite{blei2017variational, zhang2018advances}. 

In recent years, sophisticated variational algorithms have been proposed due to advances in computational power and the development of modern deep learning frameworks such as TensorFlow \cite{abadi2016tensorflow} and PyTorch \cite{paszke2019pytorch}, which enable tractable construction and learning of large scale probabilistic models. These methods either deterministically generate a set of posterior samples \cite{liu2016stein, gallego2018stochastic} or directly model a parametric probability distribution to approximate the true posterior pdf \cite{kingma2014auto, rezende2015variational, kingma2016improved, kucukelbir2017automatic}. In geophysics, bespoke variational inference methods were developed for rock physical interpretation and inversion of seismic data by \citet{nawaz2018variational, nawaz2019rapid} and \citet{nawaz2020variational}. Since then variational methods have been applied to a variety of problems including travel time tomography \cite{zhang2019seismic, zhao2021bayesian, levy2022variational}, seismic denoising \cite{siahkoohi2021preconditioned, siahkoohi2023reliable}, seismic amplitude inversion \cite{zidan2022regularized}, earthquake hypocentre inversion \cite{smith2022hyposvi}, slip distribution inversion \cite{sun2023new}, full waveform inversion in 2D \cite{zhang2020variational, urozayev2022reduced, wang2023re} and in 3D \cite{zhang20233, lomas20233d}, and survey or experimental design \cite{strutz2023variational}. In addition, various types of neural networks produce probabilistic outputs and can be considered variational methods \cite{bishop1994mixture}, and these have been applied to subsurface imaging problems for more than two decades \cite{devilee1999efficient, meier2007fully, meier2007global, ray2010efficient, shahraeeni2011fast, shahraeeni2012fast, de2013bayesian, kaufl2014framework, kaufl2016solving, earp2019probabilistic, earp2019probabilistic2, cao2020near, lubo2021exhaustive, zhao2021bayesian, zhang2021bayesian, hansen2022use, bloem2022introducing, siahkoohi2022deep}. Interestingly, \citet{zhang20233} explained how certain variational methods are related to novel Monte Carlo type algorithms, showing that a spectrum of techniques might be constructed that combine the strengths of both approaches.

The performance of variational inference methods depends on the complexity and expressiveness of the predefined variational family. There is an inherent trade-off involved in selecting a tractable set of distributions: increasing the capacity of the variational family to approximate the posterior distribution usually also increases the complexity of the optimisation problem. In many variational methods, the approximating family is fixed and constrained in ways which might exclude good approximations of the true posterior pdf. Another reason that the true posterior pdf cannot be represented accurately is that the solutions of inverse problems are often expressed using a finite basis or parametrisation, which might be chosen arbitrarily; this problem can be solved partly by trans-dimensional inversion \cite{green1995reversible, bodin2009seismic}. However, the focus of this current work is to find the true posterior solution given a predefined parametrisation. 

The mismatch between the variational family and the true posterior pdf often results in underestimation of posterior variances of the model parameters and an inability to capture posterior correlations \cite{miller2017variational}. For instance, the mean field approximation is commonly employed in variational methods in order to simplify the optimisation problem. This assumes a factorised structure for the variational distribution such as a Gaussian distribution with a diagonal covariance matrix, which ignores correlations between different parameters and can therefore yield poor inversion results \cite{bishop2006pattern, blei2017variational, zhang20233}. The trend in defining expressive variational families has mainly been to design more complex models, often using neural network based structures to achieve greater flexibility. Examples of such models include \textit{normalising flows} \cite{rezende2015variational} and their improved versions \cite{dinh2014nice, kingma2016improved, durkan2019neural, kobyzev2019normalizing, papamakarios2019normalizing}. However, building effective variational models and solving the corresponding optimisation problems, which involve a large number of parameters to be optimised, pose significant challenges.

A mixture model is a weighted sum of component probability distributions, and is useful because a general mixture model has the capability to represent any complex probability distribution to any desired level of accuracy \cite{bishop1994mixture, bishop2006pattern}. It is therefore reasonable to construct a variational family using a finite mixture of simple and parametric component distributions such as Gaussians. However, directly optimising a mixture model is a non-convex problem, so components can easily become trapped in suboptimal solutions. Additionally, it is challenging to determine the appropriate number of mixture components in advance. 

Recently, a variational method called \textit{Boosting Variational Inference} \cite[BVI --][]{guo2016boosting, miller2017variational} has been investigated. This method constructs the posterior distribution using a mixture distribution where each component of the mixture distribution is added and optimised sequentially. BVI starts by fitting a single component, then iteratively enhances the mixture model by gradually adding new component distributions. As more components are included, the posterior approximation becomes progressively more accurate, in theory thereby improving the results offered by one single component distribution. An efficient, greedy algorithm is often implemented by fixing the solution from the previous iteration, and optimising only the shape of the new component and its relative weight at each iteration. This approach avoids the need to design complex variational models a priori, but requires an additional optimisation for each added component. Similar to conventional mixture models, BVI is capable of capturing multimodality and incorporating rich covariance structures. However, unlike conventional methods, this form of BVI simplifies the objective function by focusing solely on the optimisation of a single new component at each step \cite{locatello2018boosting}. This makes the optimisation process more manageable and facilitates the construction of an expressive variational family. Since it was proposed by \citet{guo2016boosting} and \citet{miller2017variational}, various authors have contributed to the development of BVI \cite{locatello2018boosting, locatello2018boosting2, giaquinto2020gradient, campbell2019universal}. 

In this paper, our goal is to introduce BVI to non-linear geophysical inverse problems, and to solve interrogation problems efficiently using the fact that the obtained inversion results are represented by an analytic posterior expression. The paper is organised as follows. In section 2, we provide an introduction to variational Bayesian inversion and establish the BVI framework. We analyse the analytical properties of the posterior distribution and demonstrate the use of BVI and its analytical posterior expression for solving interrogation problems. In subsequent sections we apply BVI to two typical geophysical inversion problems: travel time tomography and full waveform inversion. In section 4, the analytic full waveform inversion results are used to solve an interrogation problem. Finally, we discuss our findings and draw conclusions based on our study.

\section{Methodology}
In this section, we establish methodology to answer scientific questions using interrogation theory. We show that this operation can often be summarised as evaluating the following integral with respect to a random variable $\mathbf{m}$
\begin{equation}
	\int_\mathbf{m} f(\mathbf{m})\rho(\mathbf{m})\ d\mathbf{m},
	\label{eq:integral}
\end{equation}
where $f(\mathbf{m})$ describes features of $\mathbf{m}$ that are needed to answer the question of interest, and $\rho(\mathbf{m})$ is the probability distribution function (pdf) of $\mathbf{m}$. In sections 2.1 -- 2.3, we show how \textit{boosting variational inference} can be used to calculate the probability distribution $\rho(\mathbf{m})$ given some observed data. In section 2.4, we show that the result can be used to solve an interrogation problem by estimating equation \ref{eq:integral}.

\subsection{Variational Bayesian Inversion}
Bayesian inference solves inverse problems in a probabilistic manner by evaluating the so-called \textit{posterior} pdf using Bayes' rule:
\begin{equation}
	p(\mathbf{m}|\mathbf{d}_{obs}) = \dfrac{p(\mathbf{d}_{obs}|\mathbf{m})p(\mathbf{m})}{p(\mathbf{d}_{obs})},
	\label{eq:bayes}
\end{equation}
where $p(\mathbf{m})$ is the \textit{prior} distribution of model parameters $\mathbf{m}$, which describes our knowledge about $\mathbf{m}$ prior to the inversion. The conditional probability $p(\mathbf{d}_{obs}|\mathbf{m})$ is the \textit{likelihood} of observing data $\mathbf{d}_{obs}$ given an Earth model $\mathbf{m}$. The denominator $p(\mathbf{d}_{obs}) = \int_{\mathbf{m}}{p(\mathbf{d}_{obs}|\mathbf{m}) p(\mathbf{m})d\mathbf{m}}$ is a normalisation constant called the \textit{evidence}. By combining these three terms on the right hand side, we obtain the \textit{posterior} distribution $p(\mathbf{m}|\mathbf{d}_{obs})$, which describes the probability of all possible models that are consistent with the observed data, prior information and physical forward functions used to evaluate the likelihood. Therefore, Bayesian inference provides a full inversion solution and quantifies the post inversion state of uncertainty.

Variational inference solves Bayesian problems by estimating the fixed but unknown posterior pdf. The variational goal is to select one optimal distribution $q^*(\mathbf{m})$ that best approximates the posterior pdf from a family of known distributions (called the variational family) $\mathcal{Q}(\mathbf{m})=\{q(\mathbf{m})\}$. This can be accomplished by finding the distribution $q(\mathbf{m})$ that minimises the distance (difference) between the variational and posterior distributions. The following Kullback-Leibler (KL) divergence \cite{kullback1951information} is typically used for this purpose:
\begin{equation}
	\text{KL}[q(\mathbf{m})||p(\mathbf{m}|\mathbf{d}_{obs})] = \mathbb{E}_{q(\mathbf{m})}[\log q(\mathbf{m}) - \log p(\mathbf{m}|\mathbf{d}_{obs})],
	\label{eq:kl}
\end{equation}
The KL-divergence measures the distance between two distributions $q(\mathbf{m})$ and $p(\mathbf{m}|\mathbf{d}_{obs})$. It has the property KL$[q(\mathbf{m})||p(\mathbf{m}|\mathbf{d}_{obs})] \geq 0$, with equality only when $q(\mathbf{m})=p(\mathbf{m}|\mathbf{d}_{obs})$. Evaluating the KL-divergence requires that the posterior probability $p(\mathbf{m}|\mathbf{d}_{obs})$ is calculated, which is infeasible in many problems since the evidence term $p(\mathbf{d}_{obs})$ is often analytically and computationally intractable. However, it can be shown that minimising the KL-divergence is equivalent to maximising the \textit{evidence lower bound} of $\log p(\mathbf{d}_{obs})$ (ELBO$[q(\mathbf{m})]$) defined as:
\begin{equation}
	\text{ELBO}[q(\mathbf{m})] = \mathbb{E}_{q(\mathbf{m})}[\log p(\mathbf{m}, \mathbf{d}_{obs})] - \mathbb{E}_{q(\mathbf{m})}[\log q(\mathbf{m})],
	\label{eq:elbo}
\end{equation}
This only requires that the joint probability $p(\mathbf{m}, \mathbf{d}_{obs})$ is calculated, which is computationally tractable \cite{blei2017variational}. By maximising equation \ref{eq:elbo} with respect to $q(\mathbf{m})$, we can estimate fully probabilistic solutions to Bayesian inverse problems using optimisation methods.

It is evident that the accuracy of variational inference depends on the expressiveness of the variational family $\mathcal{Q}(\mathbf{m})$. However, increasing the complexity of $\mathcal{Q}(\mathbf{m})$ to improve accuracy also tends to make the optimisation problem more challenging, or at least leads to higher-dimensional inverse problems. In the next section we will demonstrate how to mitigate this issue by employing mixtures of simpler distributions as the variational family.

\subsection{Boosting Variational Inference}
In boosting variational inference (BVI), we define the variational family to comprise the set of distributions that can be represented by a mixture of $n$ simpler component distributions
\begin{equation}
	q^n(\mathbf{m}) = \sum_{i=1}^{n} w_ig_i(\mathbf{m}),
	\label{eq:mixture}
\end{equation}
where each $g_i(\mathbf{m})$ represents a chosen mixture component. The component pdfs can theoretically be any parametric distribution (meaning that an explicit formula describes their form, with parameters that define their shape). The weight $w_i$ controls the magnitude of the contribution of each component $g_i(\mathbf{m})$, satisfying $0\le w_i \le 1$ and $\sum_{i=1}^{n} w_i = 1$. Remarkably, the mixture in equation \ref{eq:mixture} can approximate any target distribution to any level of accuracy, even when using a simple base distribution $g_i(\mathbf{m})$ \cite{bishop1994mixture, meier2007global, shahraeeni2011fast, earp2019probabilistic, earp2019probabilistic2}. 

Directly maximising ELBO$[q^n(\mathbf{m})]$ with respect to the variational parameters $\{w_i, g_i(\mathbf{m}); i = 1,2,...,n\}$ is a non-convex problem so algorithms may converge to local maxima at which one component dominates while the weights of other components become negligible \cite{guo2016boosting}. The gradient boosting approach \cite{friedman2001greedy, meir2003introduction} can be used to solve this problem. The main idea is to sequentially add components to an ensemble, each being used to correct errors of its predecessors. Inspired by this, we determine an optimal variational distribution $q_n(\mathbf{m})$ through an iterative procedure, adding one new component distribution to the mixture model at each step. The procedure begins with a single component $q^1(\mathbf{m}) = g_1(\mathbf{m})$ with $w_1 = 1$. We fit $g_1(\mathbf{m})$ using a traditional variational objective function \cite{blei2017variational, zhang2018advances}. Note that in a linear problem optimising a single component (e.g., a single Gaussian distribution) by maximising ELBO$[q^1(\mathbf{m})]$ gives precisely the Bayesian least squares solution \cite{tarantola1982generalized, valentine2023emerging}. In each subsequent step $t=2,3,...,n$, BVI adds one new component $g_t$ to the mixture model, with weight $w_t \in [0, 1]$. The new distribution $q^t(\mathbf{m})$ is constructed by combining the previous mixture distribution $q^{t-1}(\mathbf{m})$, weighted by $(1-w_t)$, with the new component $g_t(\mathbf{m})$ weighted by $w_t$:
\begin{equation}
	q^{t}(\mathbf{m}) = (1-w_t)q^{t-1}(\mathbf{m}) + w_tg_{t}(\mathbf{m}),
	\label{eq:mix_2pdf}
\end{equation}
We then maximise ELBO$[q^t(\mathbf{m})]$ with respect to $w_t$ and $g_t$.

Since jointly optimising $w_t$ and $g_t(\mathbf{m})$ is also a non-convex problem, we adopt a sequential approach which finds the optimal component $g_t(\mathbf{m})$ first, then finds the corresponding weight $w_t$ \cite{locatello2018boosting}. Based on equation \ref{eq:mix_2pdf}, we treat the new mixture pdf $q^{t}(\mathbf{m})$ as a perturbation from the current distribution $q^{t-1}(\mathbf{m})$, where the component distribution $g_{t}(\mathbf{m})$ describes the shape of the perturbation and $w_t \in [0, 1]$ describes the size of the perturbation. We take the first-order Taylor expansion of ELBO$[q^t(\mathbf{m})]$ around $q^{t-1}(\mathbf{m})$ \cite{locatello2018boosting2}:
\begin{equation}
	\medmath{
		\begin{split}
			\text{ELBO}[q^{t}(\mathbf{m})] &= \text{ELBO}[q^{t-1}(\mathbf{m}) + w_tg_{t}(\mathbf{m}) - w_tq^{t-1}(\mathbf{m})] \\
			& = \text{ELBO}[q^{t-1}(\mathbf{m})] + w_t \left<g_{t}(\mathbf{m}), \nabla\text{ELBO}[q^{t-1}(\mathbf{m})]\right> - w_t \left<q^{t-1}(\mathbf{m}), \nabla\text{ELBO}[q^{t-1}(\mathbf{m})]\right> + o(w_t^2),
	\end{split}}
	\label{eq:taylor_elbo}
\end{equation}
where $\left<x(\theta), y(\theta)\right> = \int x(\theta)y(\theta)d\theta$ calculates the inner product between functions $x(\theta)$ and $y(\theta)$. The term $\nabla\text{ELBO}[q^{t-1}(\mathbf{m})] = \log \dfrac{p(\mathbf{m}, \mathbf{d}_{obs})}{q^{t-1}(\mathbf{m})}$ is the functional gradient of the ELBO with respect to $q^{t-1}(\mathbf{m})$. In order to maximise ELBO$[q^{t}(\mathbf{m})]$ in equation \ref{eq:taylor_elbo}, we must choose the $g_{t}(\mathbf{m})$ that maximises $\left<g_{t}(\mathbf{m}), \nabla\text{ELBO}[q^{t-1}(\mathbf{m})]\right>$ since $q^{t-1}(\mathbf{m})$ is fixed:
\begin{equation}
	g_t(\mathbf{m}) = \argmax_{g_t(\mathbf{m})} \left<g_{t}(\mathbf{m}), \nabla\text{ELBO}[q^{t-1}(\mathbf{m})]\right> = \argmax_{g_t(\mathbf{m})} \left<g_t(\mathbf{m}), \log \dfrac{p(\mathbf{m}, \mathbf{d}_{obs})}{q^{t-1}(\mathbf{m})}\right>,
	\label{eq:optimal_gt}
\end{equation}
Direct maximisation of the inner product in equation \ref{eq:optimal_gt} is ill-posed and can lead to $g_t(\mathbf{m})$ degenerating into a narrow distribution or even a single point mass which only has non-zero probability value at the maximum of $\nabla\text{ELBO}[q^{t-1}(\mathbf{m})]$ -- a degenerate probability distribution that has zero width. To solve this problem, we introduce an additional regularisation term that involves the entropy of $g_t(\mathbf{m})$, given by the negative scalar product of $g_t(\mathbf{m})$ and $\log g_t(\mathbf{m})$ \cite{guo2016boosting}:
\begin{equation}
	\begin{split}
		g_t(\mathbf{m}) & = \argmax_{g_t(\mathbf{m})} \left<g_t(\mathbf{m}), \nabla \text{ELBO}[q^{t-1}(\mathbf{m})]\right> - \lambda \left<g_t(\mathbf{m}), \log g_t(\mathbf{m})\right> \\
		& = \argmax_{g_t(\mathbf{m})} \mathbb{E}_{g_t(\mathbf{m})}[\log p(\mathbf{m}, \mathbf{d}_{obs})] - \mathbb{E}_{g_t(\mathbf{m})}[\log q^{t-1}(\mathbf{m})] - \lambda \mathbb{E}_{g_t(\mathbf{m})}[\log g_t(\mathbf{m})],
	\end{split}
	\label{eq:optimal_gt_reg}
\end{equation}
where $\mathbb{E}_{g_t(\mathbf{m})}[\cdot]$ calculates the expectation of any function with respect to $g_t(\mathbf{m})$. Parameter $\lambda$ is a regularisation factor that controls the weight of the entropy term. Entropy measures the uncertainty represented by any pdf, so by maximising the entropy we ensure that the pdf does not collapse to a narrow, effectively degenerate distribution. \citet{locatello2018boosting2} referred to the objective function in equation \ref{eq:optimal_gt_reg} as the \textit{residual evidence lower bound} (RELBO$[g_t(\mathbf{m})]$)
\begin{equation}
	\text{RELBO}[g_t(\mathbf{m})] := \mathbb{E}_{g_t(\mathbf{m})}[\log p(\mathbf{m}, \mathbf{d}_{obs})] - \mathbb{E}_{g_t(\mathbf{m})}[\log q^{t-1}(\mathbf{m})] - \lambda \mathbb{E}_{g_t(\mathbf{m})}[\log g_t(\mathbf{m})],
	\label{eq:relbo}
\end{equation}
The expectation terms and their gradients in both equations \ref{eq:elbo} and \ref{eq:relbo} can be estimated using Monte Carlo integration \cite[details can be found in][]{zhao2021bayesian}. Since we would normally perform many iterations to maximise these two equations, we can use a relatively small number of samples per iteration \cite[even only a single sample --][]{kucukelbir2017automatic}. By maximising this objective function, we can find an optimal $g_t(\mathbf{m})$ at each step of the algorithm.

In equation \ref{eq:optimal_gt}, $\log \frac{p(\mathbf{m}, \mathbf{d}_{obs})}{q^{t-1}(\mathbf{m})}$ describes the residual discrepancy between the current variational distribution $q^{t-1}(\mathbf{m})$ and the joint probability distribution $p(\mathbf{m}, \mathbf{d}_{obs}) = p(\mathbf{d}_{obs}) p(\mathbf{m}|\mathbf{d}_{obs})$ which is equal to the unnormalised posterior distribution  $p(\mathbf{m}|\mathbf{d}_{obs})$ according to equation \ref{eq:bayes}. If $q^{t-1}(\mathbf{m})$ is proportional to the true (normalised) posterior pdf, i.e., $q^{t-1}(\mathbf{m}) \propto p(\mathbf{m}|\mathbf{d}_{obs})$ everywhere in the parameter space, the above residual would be constant. However, in most situations this residual has peaks where the current variational distribution underestimates the posterior distribution, and has basins where $q^{t-1}(\mathbf{m})$ overestimates $p(\mathbf{m}|\mathbf{d}_{obs})$. By introducing a new component $g_t(\mathbf{m})$, we aim to add density to regions where $q^{t-1}(\mathbf{m})$ underestimates and (through the relative weighting scheme in equation 5) weaken regions where it overestimates the posterior pdf (this can be proven using information theory \cite{jaynes1957information}). The goal is to find an optimal $g_t(\mathbf{m})$ that maximises $\large(\mathbb{E}_{g_t(\mathbf{m})}[\log p(\mathbf{m}, \mathbf{d}_{obs})] - \mathbb{E}_{g_t(\mathbf{m})}[\log q^{t-1}(\mathbf{m})]\large)$, which can be interpreted as minimising the cross entropy of $g_t(\mathbf{m})$ with respect to $p(\mathbf{m}, \mathbf{d}_{obs})$ and maximising that with respect to $q^{t-1}(\mathbf{m})$. In other words, $g_t(\mathbf{m})$ should be as close as possible to the (unnormalised) posterior distribution, and at the same time should be different from the current approximation -- it should capture the aspects of the posterior pdf that the current mixture distribution cannot yet approximate. This allows BVI to gradually improve the accuracy of the variational distribution by iteratively adding new components.

There are three commonly used methods to determine the weight coefficient $w_t \in [0,1]$ for the new component in BVI. The first method uses an empirical formula to guarantee a series of decreasing weights for each additional component \cite{locatello2018boosting2, locatello2018boosting}:
\begin{equation}
	w_t = \dfrac{2}{t+1}, \quad t = 1,2,...,n,
	\label{eq:weight_1}
\end{equation}
Although this formula abandons the ideal of finding optimal weight coefficients, it provides a straightforward approach to update the weight. Any error caused by non-optimality of this scheme can be corrected by the introduction of additional components to the mixture distribution.

The second method for updating weight coefficients involves a line search. The weight is updated by maximising ELBO$[q^t(\mathbf{m})]$ (note this is not maximising RELBO$[g_t(\mathbf{m})]$ with respect to $w_t$) \cite{guo2016boosting}:
\begin{equation}
	w_t^{(k+1)} = w_t^{(k)} + \dfrac{b}{k}\nabla_{w_t} \text{ELBO}[q^t(\mathbf{m})],
	\label{eq:weight_2}
\end{equation}
where superscripts (k+1) and (k) represent two consecutive iterations, and $b$ is the initial step size decayed by $1/k$. The method to calculate $\nabla_{w_t} \text{ELBO}[q^t(\mathbf{m})]$ is provided in \ref{ap:A}.

The third method, updates the weights for all components when each new component is added to the mixture model \cite{locatello2018boosting2}:
\begin{equation}
	\mathbf{w}^{(k+1)} = \mathbf{w}^{(k)} + \dfrac{b}{k}\nabla_{\mathbf{w}} \text{ELBO}[q^t(\mathbf{m})],
	\label{eq:weight_3}
\end{equation}
where $\mathbf{w} = [w_1, w_2,..., w_t]^T$ is a vector containing the weights of all components. The gradient term can be calculated similarly to the line search method (\ref{ap:A}).

Once the weight coefficient is obtained the new mixture distribution $q^t(\mathbf{m})$ can be constructed by combining the new component $g_t(\mathbf{m})$ with the existing mixture distribution using Equation \ref{eq:mix_2pdf}.

\subsection{BVI using Gaussian Components}
In this work, we use Gaussian component distributions as the mixture components: $g_i(\mathbf{m}) = \mathcal{N}(\mathbf{m};\mu_i, \Sigma_i)$, parametrised by a mean vector $\mu_i$ and a covariance matrix $\Sigma_i$. A mixture of Gaussians is capable of representing any target distributions to any accuracy \cite{bishop2006pattern}. For each component, we optimise $\mu$ and $\Sigma$ by maximising the RELBO in equation \ref{eq:relbo}, and below we test the three schemes to determine the weights. Once convergence is achieved, the obtained Gaussian component is added to form the new mixture distribution. 

Considering that model parameters in many geophysical inverse problems are subject to hard constraints (e.g., seismic velocity must be greater than zero), and Gaussian distributions and their mixtures are defined in the unbounded space of Real numbers, we apply the inverse logistic function to transform the mixture distribution from the space of Real numbers into the constrained space \cite{zhang2019seismic}. This transformation is defined as:
\begin{equation}
	\left\{
	\begin{array}{rl}
		\mathbf{m}&= f(\mathbf{z}) = \mathbf{a} + \dfrac{\mathbf{b}-\mathbf{a}}{1+\exp (-\mathbf{z})},\\
		\log p(\mathbf{m}|\mathbf{d}_{obs})&= \log p(\mathbf{z}) - \log|\det \partial_{\mathbf{z}} f(\mathbf{z})| \\
		&= \log \sum_{i}w_i \mathcal{N}(\mathbf{z};\mu_i, \Sigma_i) - \log|\det \partial_{\mathbf{z}} f(\mathbf{z})|,
	\end{array}
	\right.
	\label{eq:log_transform}
\end{equation}
where $\mathbf{m}$ and $\mathbf{z}$ are model parameters in the constrained and unconstrained spaces, respectively. Hyper-parameters $\mathbf{a}$ and $\mathbf{b}$ are lower and upper bounds on $\mathbf{m}$, and are fixed during the optimisation. In the second equation, $p(\mathbf{z})$ is the mixture distribution obtained using BVI in the space of Real numbers. The term $|\det(\cdot)|$ calculates the absolute value of the determinant of the Jacobian matrix corresponding to this transform, which accounts for the volume change. We use equation \ref{eq:log_transform} to transform each parameter in vector $\mathbf{z}$ to that in $\mathbf{m}$, such that the corresponding Jacobian matrix is a diagonal matrix and its determinant is analytic and easy to calculate. This means that the correlation information of vector $\mathbf{m}$ is purely determined by the covariance matrices $\Sigma_i$ (therefore we do not lose posterior correlations by applying this transform). As a result, the posterior distribution modelled using the proposed BVI algorithm, as well as its statistical properties, can be represented analytically. 

If only a single Gaussian component is involved in equation \ref{eq:log_transform}, BVI becomes automatic differentiation variational inference \cite[ADVI -- ][]{kucukelbir2017automatic} -- another well-established variational method that tries to fit (approximate) the true posterior pdf with a Gaussian distribution \cite{tarantola1982generalized, valentine2023emerging}.
ADVI also provides an analytic approximation to the posterior distribution, and usually seems to estimate the mean model accurately. However, due to its theoretical assumption of a single Gaussian distribution in the unconstrained space, the method usually underestimates parametric uncertainty around the mean. By adding more Gaussian components we regard BVI as an iterative method to enhance the performance of ADVI.

\begin{figure}
	\centering\includegraphics[width=0.6\textwidth]{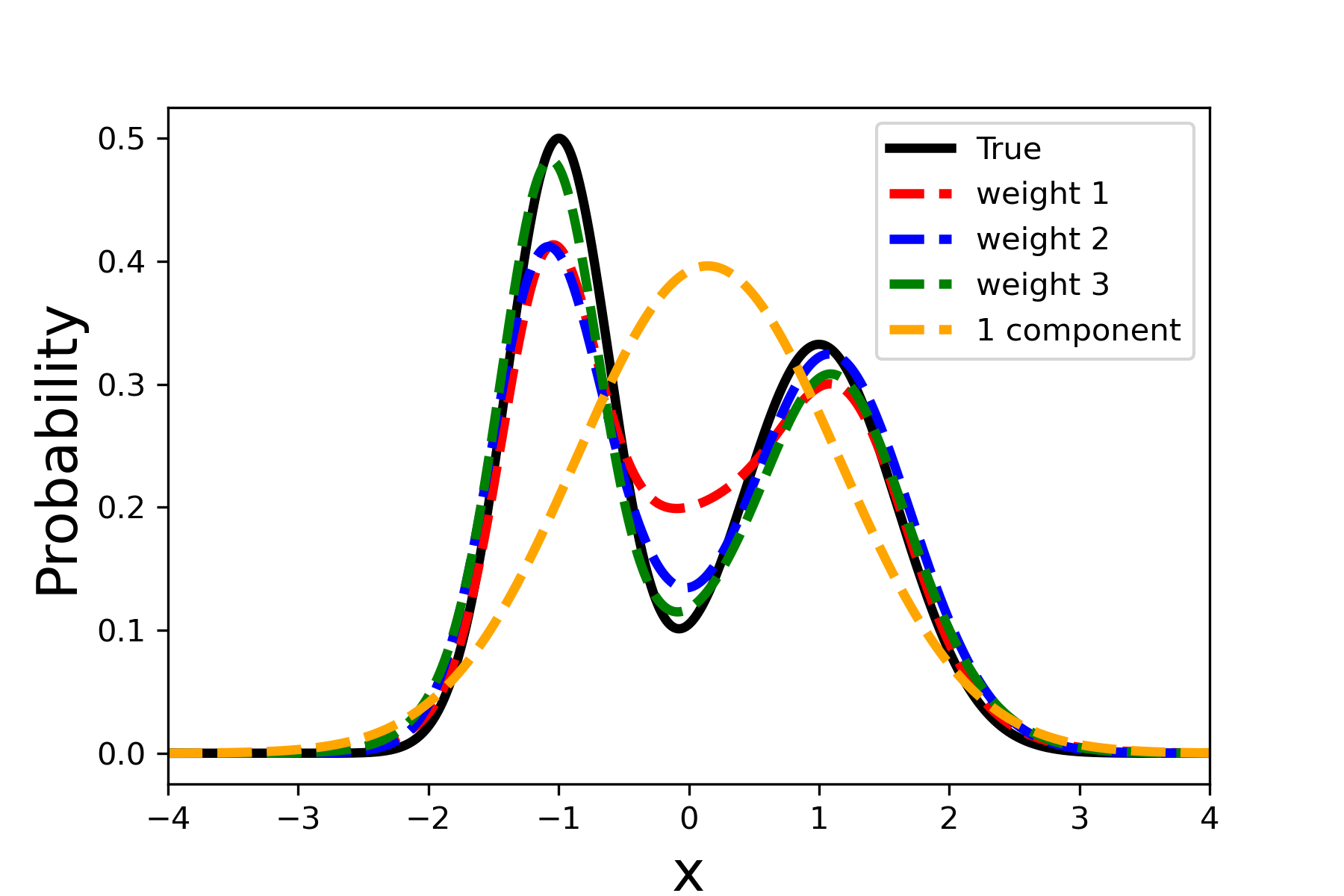}
	\caption{BVI results obtained using 3 different weight calculation methods. Black line represents the target distribution, while the dashed orange line shows the result from conventional variational inference without boosting, which uses a single Gaussian component (essentially the ADVI method). Dashed red, blue, and green lines correspond to the results obtained using different weight calculation methods in equations \ref{eq:weight_1}, \ref{eq:weight_2} and \ref{eq:weight_3}. The last two methods yield better results but require additional computations.}
	\label{fig:BVI_1D_compare}
\end{figure}

Figure \ref{fig:BVI_1D_compare} shows a toy example that demonstrates the performance of BVI with Gaussian components. The target posterior distribution is a mixture of two Gaussian distributions: $p(x) = 0.5\mathcal{N}(x;-1, 0.4) + 0.5\mathcal{N}(x;1, 0.6)$, represented by the black line in Figure \ref{fig:BVI_1D_compare}. To apply BVI, we first optimise the initial component by maximising the ELBO in equation \ref{eq:elbo}, which is equivalent to a conventional variational problem. The dashed orange line in Figure \ref{fig:BVI_1D_compare} shows the first component after convergence. It is evident that this single Gaussian distribution fails to approximate the bimodal posterior distribution accurately, highlighting the limitations of ADVI, and variational methods in general when an inappropriate variational family that does not include the true posterior pdf is chosen.

We then iteratively add more Gaussian components to the mixture model by maximising the RELBO using equation \ref{eq:relbo}. We compare the performance of the 3 different weight calculation methods in equations \ref{eq:weight_1}, \ref{eq:weight_2} and \ref{eq:weight_3}. In each test, we boost the posterior distribution by adding 40 Gaussian components so as effectively to ensure full convergence of BVI. The reason that a large number of components is used is partly because the first component, which achieves a poor approximation, is fixed thus more components need to be introduced to correct its error. In addition, since we generally do not know the true posterior distribution, we do not know when to stop the algorithm unless convergence is observed. The results using equations \ref{eq:weight_1}, \ref{eq:weight_2} and \ref{eq:weight_3} are shown by the dashed red, blue and green lines in Figure \ref{fig:BVI_1D_compare}, respectively. All three methods provide a fair approximation to the true posterior distribution, with the first method performing the worst and the third method performing the best. However, the second and third methods require additional computations to estimate the gradient of the ELBO in equations \ref{eq:weight_2} and \ref{eq:weight_3}, which involve evaluating the posterior distribution many times. This example demonstrates that even the simple fixed weight method significantly improves upon the initial variational solution (dashed orange line in Figure \ref{fig:BVI_1D_compare}) without any additional computational complexity. Since we are interested in applying these methods to high-dimensional problems, minimising computational complexity is paramount if we are to find meaningful solutions. In the subsequent inversion examples, we therefore employ the fixed weight calculation method, but highlight that in other circumstances practitioners might prefer a different choice.

\subsection{Probabilistic Interrogation using BVI}
In interrogation theory we try to find the optimal answer to a scientific question, where a utility function $U(a|\mathbf{m})$ is defined which quantifies the net benefits associated with accepting any possible answer $a$ given that model $\mathbf{m}$ is true. The optimal answer $a^*$ is found by maximising this utility function within the space of possible answer: $a^* = \underset{a\in \mathbb{A}}{\arg\max} \ U(a)$. To reduce the complexity of this maximisation problem, \citet{arnold2018interrogation} introduced a target space $\mathbb{T}$ such that the scientific question $Q$ can be answered directly within this space. They defined a target function $T(\mathbf{m})$ that maps the high dimensional model parameter $\mathbf{m}$ to a low dimensional target space parameter values $t$. A simplified utility function can then be defined as $U(a|t)$. One of the utility functions considered in \citet{arnold2018interrogation} is a negative squared error function:
\begin{equation}
	U(a|t) = U(a|t) = -(a-t)^2,
	\label{eq:utility}
\end{equation}
in which $t$ is considered to represent the true state in the target space. 
This formulation leads to an analytical expression for the optimal (minimum bias) answer:
\begin{equation}
	a^* = \mathbb{E}[T(\mathbf{m})|\mathbf{d}_{obs}] = \int_\mathbf{m} T(\mathbf{m})p(\mathbf{m}|\mathbf{d}_{obs})\ d\mathbf{m},
	\label{eq:optimal_answer}
\end{equation}
Equation \ref{eq:optimal_answer} states that the optimal answer corresponds to the posterior expectation of the target function, and different forms for this expression result from different choices of utility function in equation \ref{eq:utility} \cite{arnold2018interrogation}.

In previous works \cite{zhao2022interrogating, zhang2021interrogation}, the target function was assumed to be deterministic, meaning that the target value was uniquely determined given a model sample $\mathbf{m}$. Consequently, uncertainty in the answer was attributed solely to uncertainty in the inversion process. In reality, the definition of the target function often incorporates knowledge from a variety of experts, which introduces human biases and uncertainties \cite{o2006uncertain, polson2010dynamics, bond2012makes}. In an interrogation example below, we show that biased judgments from different individuals can lead to incorrect answers. To address the uncertainty in the final answer caused by the deterministic target function in order to mitigate bias, we use fully probabilistic target functions.

Define a random variable $\tau$ to represent different states of possible target function values, with an associated probability distribution function $p(\tau)$. This approach characterizes the nondeterministic behaviour of the target function and addresses the inherent uncertainty. The optimal answer, which calculates the posterior mean of the summarized state $\tau$, is given by
\begin{equation}
	\begin{split}
		a^* = \mathbb{E}[\tau|\mathbf{d}_{obs}] & = \int_\mathbf{m} \int_{\tau} \tau p(\tau, \mathbf{m}|\mathbf{d}_{obs})\ d\mathbf{m} d\tau \\
		& = \int_\mathbf{m} \int_{\tau} \tau p(\tau|\mathbf{m}, \mathbf{d}_{obs}) p(\mathbf{m}|\mathbf{d}_{obs}) \ d\mathbf{m} d\tau \\
		& = \int_\mathbf{m} \int_{\tau} \tau p(\tau|\mathbf{m}) d\tau p(\mathbf{m}|\mathbf{d}_{obs}) \ d\mathbf{m},
	\end{split}
	\label{eq:probabilistic_interrogation}
\end{equation}
In the first line, we extend the deterministic target function from equation \ref{eq:optimal_answer} to a probabilistic formulation using the law of total probability $p(x) = \int_{y}p(x,y)dy$. Following \citet{siahkoohi2022deep}, we assume that the target function value $\tau$ and the observed data $\mathbf{d}_{obs}$ are conditionally independent given the model parameter $\mathbf{m}$, when using interrogation theory to solve a decision problem that maps specific information from the inversion results. This assumption leads to the third line in equation \ref{eq:probabilistic_interrogation}. The inner integral $\mathbb{E}[\tau|\mathbf{m}] := \int_{\tau} \tau p(\tau|\mathbf{m}) d\tau$ captures uncertainty in the target function value which represents the uncertainty in the interrogation process, while the outer integral accounts for uncertainty in the inversion process. Note that the above conditional independence assumption does not hold when solving a design problem using interrogation theory, as the optimal answer, which is the best design in this context, depends on the different datasets that would be observed given any considered design \cite{arnold2018interrogation, strutz2023variational}.

To summarise, equation \ref{eq:probabilistic_interrogation} can be viewed as a more general version of the original interrogation framework, achieved by considering a random variable $\tau$ with a probability distribution function $p(\tau)$ which allows for the incorporation of uncertainty in the target function. When $p(\tau)$ is defined as a Dirac delta function, denoted by $p(\tau) = \delta_{(\tau =T)}(\tau)$, where $T$ represents the deterministic target function in equation \ref{eq:optimal_answer}, equation \ref{eq:probabilistic_interrogation} reduces to equation \ref{eq:optimal_answer}. Thus, equation \ref{eq:probabilistic_interrogation} encompasses the original framework as a special case when the target function is deterministic.

Monte Carlo integration can be used to evaluate equation \ref{eq:probabilistic_interrogation}. First, we draw random model samples from the posterior distribution $p(\mathbf{m}|\mathbf{d}_{obs})$. Given each posterior sample, we generate an ensemble of possible target function values from $p(\tau|\mathbf{m})$. By combining these target values, the posterior distribution of the answer $a$ can be obtained, and the optimal answer $a^*$ to the question $Q$ is the expectation of this distribution.

In the previous sections we showed that BVI provides an analytic expression of the posterior distribution. Directly incorporating this analytic result into equations above using either the deterministic or probabilistic target function is unfortunately non-trivial because the definition of the target function often contains some conceptual process which is easier to evaluate using posterior samples and is difficult to formulate as an explicit expression. In an interrogation example provided below, the calculation of the target function requires the largest continuous body within a velocity model to be found, which is not straightforward to perform using the analytic posterior expression. To address this, we propose an implicit approach. In the BVI framework the posterior distribution is approximated in the Real (unconstrained) space as a mixture of Gaussian distributions, and significant information is captured by the mean vectors $\mu_i$ of the set of components. We transform these mean vectors $\mu_i$ back to the constrained space using equation \ref{eq:log_transform} after which the transformed vectors $\mathbf{m}_i$ can be treated as a set of representative samples, weighted by the coefficient $w_i$ corresponding to each Gaussian component in BVI. We use these samples to partly represent the full posterior pdf, and the optimal answer in equations \ref{eq:optimal_answer} and \ref{eq:probabilistic_interrogation} can be approximated as 
\begin{equation}
	a^* = \int_\mathbf{m} T(\mathbf{m})p(\mathbf{m}|\mathbf{d}_{obs})\ d\mathbf{m} \approx \sum_{i}w_i T(\mathbf{m_i}),
	\label{eq:optimal_answer_bvi}
\end{equation}
for the deterministic case, and 
\begin{equation}
	\begin{split}
		a^* &= \int_\mathbf{m} \int_{\tau} \tau p(\tau|\mathbf{m}) d\tau p(\mathbf{m}|\mathbf{d}_{obs}) \ d\mathbf{m} \\
		&\approx \sum_{i}w_i \int_{\tau} \tau p(\tau|\mathbf{m}_i) d\tau = \sum_{i}w_i \mathbb{E}[\tau|\mathbf{m}_i],
	\end{split}
	\label{eq:probabilistic_optimal_answer_bvi}
\end{equation}
for the probabilistic case. Since only tens of components are used in BVI to approximate the posterior distribution, the target function is calculated using the same number of samples from BVI. This computational simplicity is particularly important when the target function itself is computationally expensive to evaluate, especially in the case of interrogation using probabilistic target functions, and as we show below, equations \ref{eq:optimal_answer_bvi} and \ref{eq:probabilistic_optimal_answer_bvi} can still enable accurate interrogation.

\section{Travel Time Tomography}
Seismic travel time tomography is a typical non-linear geophysical inverse problem used to image the Earth's interior. The underground seismic velocity structure is mapped using measured first-arrival travel times of waves travelling between source and receiver locations. In this section, we present two tomographic examples to demonstrate the performance of BVI.

\subsection{Synthetic Example}
\begin{figure}
	\centering\includegraphics[width=0.4\textwidth]{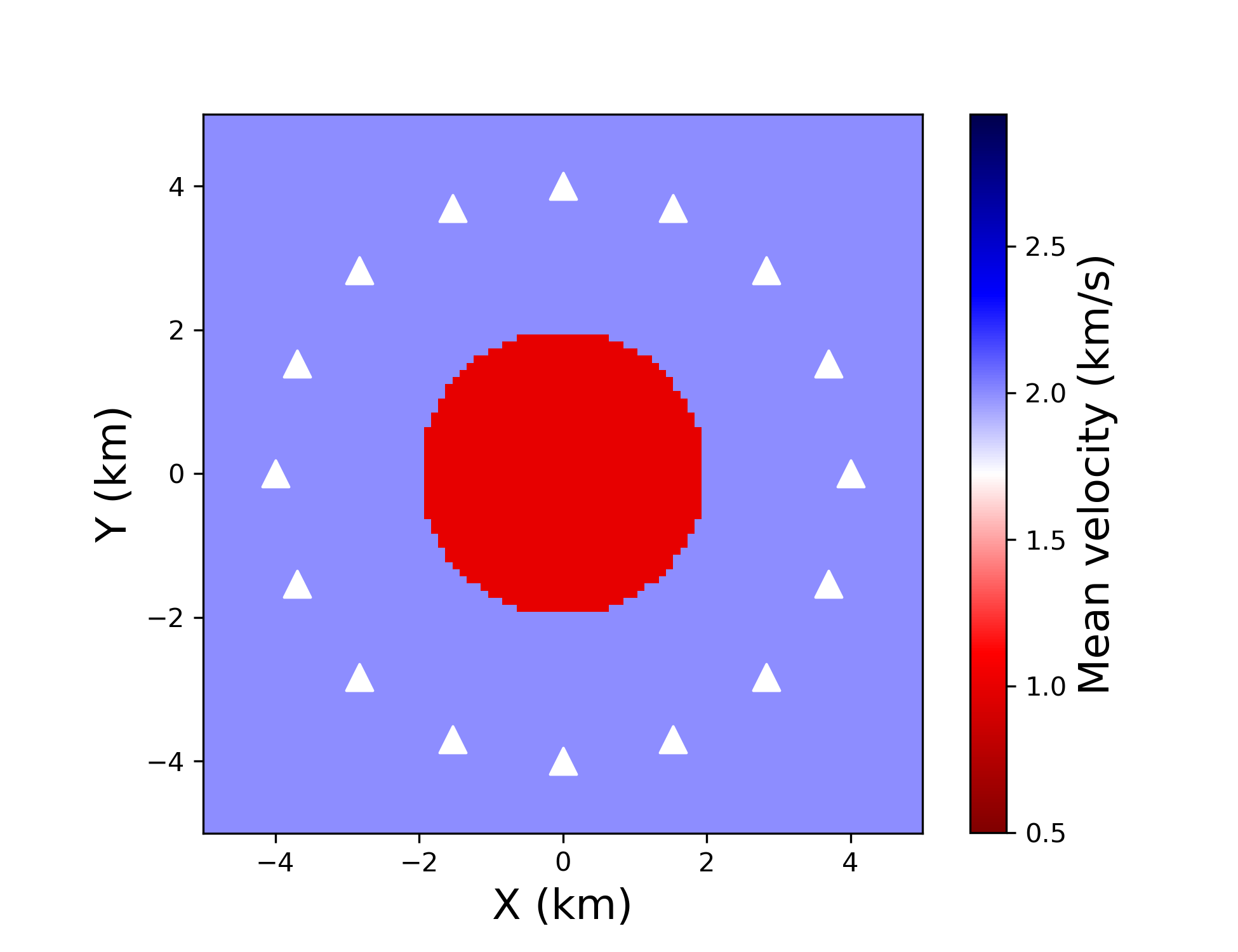}
	\caption{True velocity model of the 2D synthetic test. A low velocity circular anomaly with velocity 1 $km/s$ is embedded within a background velocity of 2 $km/s$. White triangles show locations of 16 receivers (and sources), and travel times between each pair of locations form the observed data set in this example.}
	\label{fig:low_vel_anomaly}
\end{figure}

The first example is a 2D synthetic test. Figure \ref{fig:low_vel_anomaly} shows the true velocity model, which consists of a circular low velocity anomaly of 1 km/s surrounded by a high velocity background of 2 km/s. White triangles show the locations of 16 receivers, and we assume that each receiver can also be used as a virtual source using seismic interferometry \cite{campillo2003long, wapenaar2004retrieving, curtis2006seismic}. 120 inter-receiver first-arrival travel times of waves that travel between each pair of receiver locations form the data set for this problem. For inversion we parametrise the model parameter $\mathbf{m}$ (the velocity vector) into 21 $\times$ 21 regular grid cells with a grid size of 0.5 km in both directions. We define a Uniform prior probability distribution bounded between 0.5 and 3.0 km/s for each grid cell. The likelihood function is assumed to be a diagonal Gaussian distribution with a data uncertainty $\sigma_d = 0.05$ s for all data points. We solve the forward problem to predict synthetic data using the fast marching method \cite[FMM -- ][]{rawlinson2005fast}. 

\begin{figure}
	\centering\includegraphics[width=\textwidth]{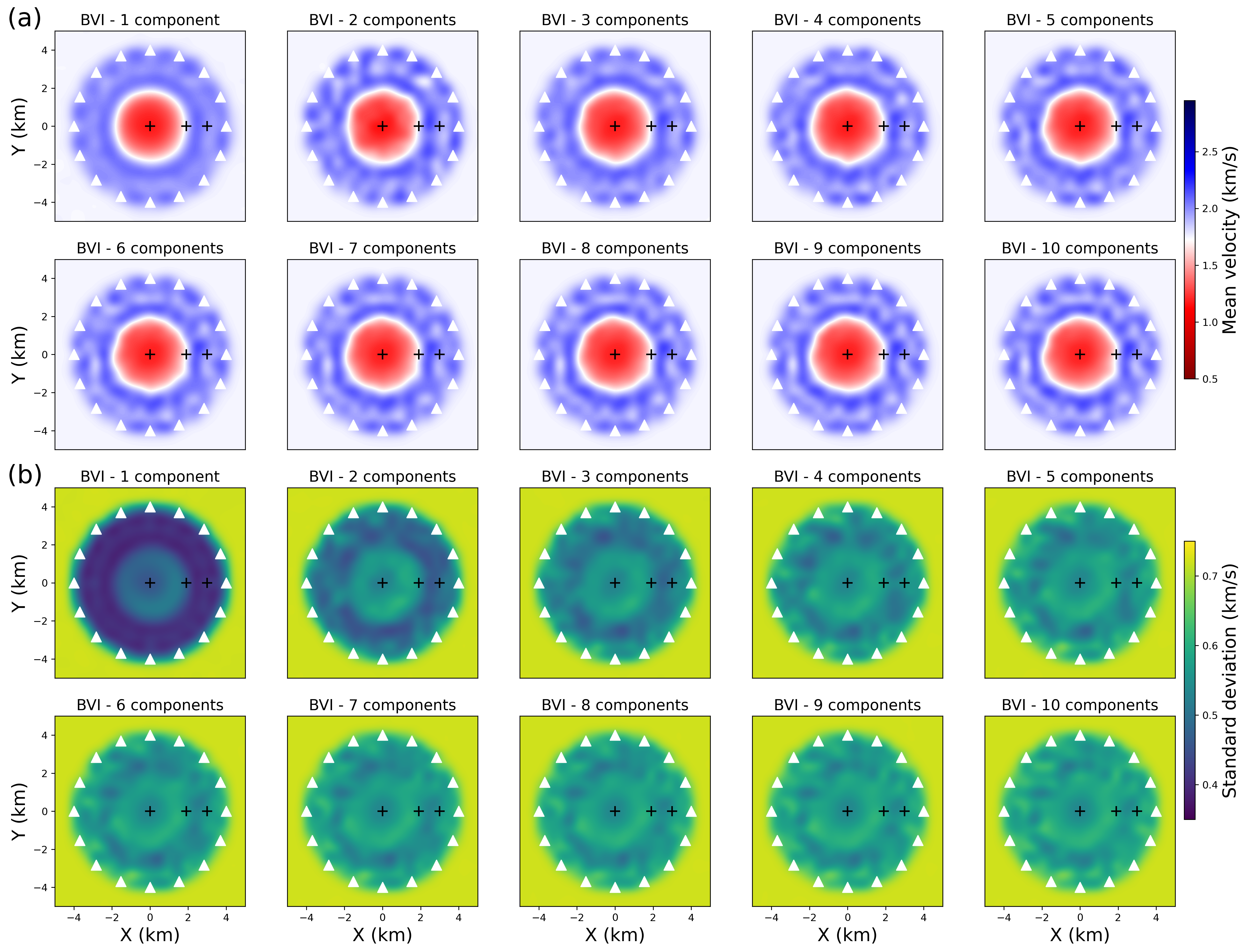}
	\caption{(a) Mean and (b) standard deviation maps of the posterior distribution obtained using BVI with different number of Gaussian components denoted in the title of each subfigure. White triangles show the 16 receiver locations and black crosses denote three specific locations whose marginal distributions are compared in Figure \ref{fig:tomo_low_vel_marginal}.}
	\label{fig:tomo_low_vel_mean_std}
\end{figure}

For BVI we use a diagonal Gaussian distribution for all component distribution, and the empirical formula in equation \ref{eq:weight_1} to calculate weight coefficients. The first component is obtained by maximising the ELBO in equation \ref{eq:elbo} which is equivalent to mean field ADVI \cite{kucukelbir2017automatic}. In subsequent BVI iterations, we sequentially optimise new components by maximising the RELBO in equation \ref{eq:relbo}. We combine the obtained Gaussian components into a mixture distribution and transform it back to the constrained space using equation \ref{eq:log_transform}. The resulting distribution is used to approximate the true posterior distribution. For each component, we update the diagonal Gaussian distribution for 5000 iterations. Within each iteration, we draw 2 random samples from the (current) variational distribution $q(\mathbf{m})$, calculate the forward function values of these samples (to get likelihood and posterior values), and use them to approximate the RELBO (or ELBO for the first component) and its gradient using Monte Carlo integration. Note that since we would normally update RELBO (or ELBO) with many iterations, we can use a relatively small number of samples (e.g., 2 samples in this example) per iteration for the numerical integration, and the overall number of samples used to approximation the integration remains large. Even though the estimated value in each iteration may then be inaccurate, the mean value over many iterations should be approximately correct \cite{kingma2014auto, zhao2021bayesian}. To test the convergence performance of BVI, we greedily add 10 components by which point the statistics of the posterior pdf show no substantial change with each iteration, as shown in Figure \ref{fig:tomo_low_vel_mean_std}.

Figures \ref{fig:tomo_low_vel_mean_std}a and \ref{fig:tomo_low_vel_mean_std}b show the mean and standard deviation maps of the posterior distribution obtained using BVI with different Gaussian components. All of these maps are calculated analytically from the BVI solution without drawing any posterior samples, using equation \ref{eq:log_transform}. Within the receiver array, the mean models effectively recover the circular low velocity anomaly and are similar to the true velocity model shown in Figure \ref{fig:low_vel_anomaly} even with only 1 component which corresponds to mean field ADVI as discussed previously. However, as expected the uncertainty map obtained using one component significantly underestimates uncertainties. As we introduce more components, the posterior uncertainties increase. The mean and standard deviation maps essentially converge such that no significant changes are observed after adding 6 -- 7 components. We observe two higher uncertainty loops in the uncertainty maps: the inner one is located at the boundary of the low velocity anomaly and arises from variations in anomaly shapes and velocity values among the plausible models that fit the observed data, and the other loop corresponds to the lower average velocity loop between the receiver array and the central anomaly, potentially because the observed data exhibits lower sensitivity in this region, as observed in previous studies \cite{galetti2015uncertainty, zhang2019seismic, zhao2021bayesian}.

\begin{figure}
	\centering\includegraphics[width=0.8\textwidth]{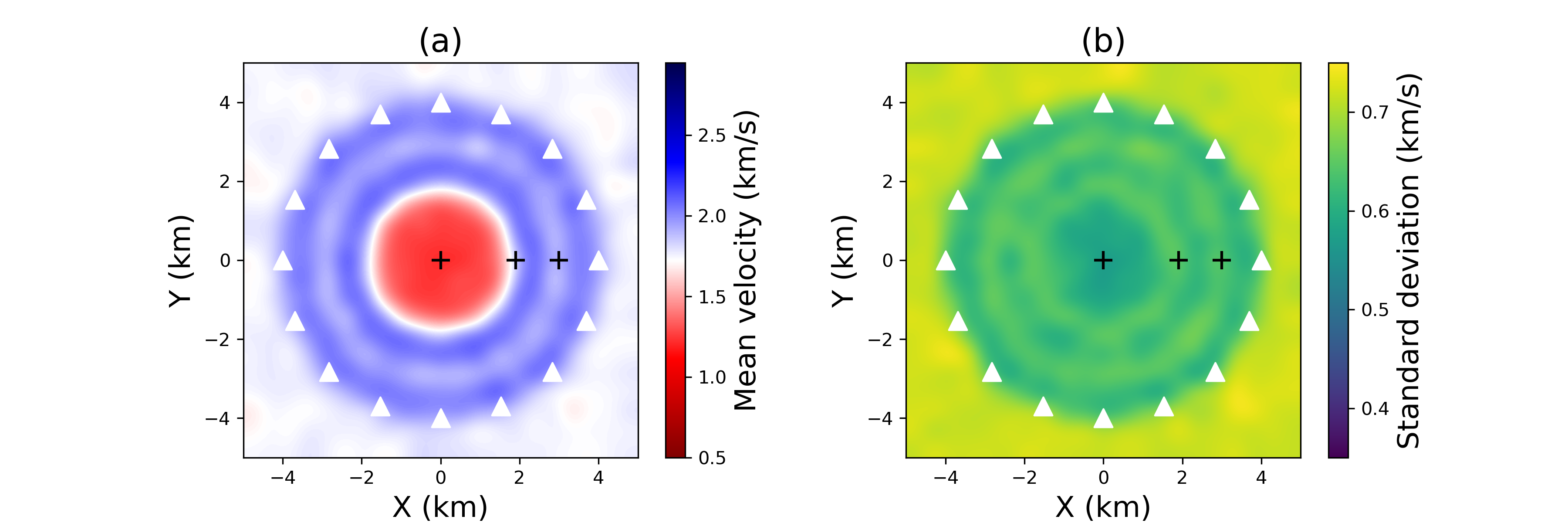}
	\caption{(a) Mean and (b) standard deviation maps obtained using MH-McMC. This result serves as the reference solution for this Bayesian tomographic problem.}
	\label{fig:tomo_low_vel_mcmc}
\end{figure}

Metropolis-Hastings Markov chain Monte Carlo (MH-McMC) was also run to estimate the solution for comparison. We ran 12 chains in parallel, each drawing 1 million samples to ensure convergence. After sampling, we discard the first 500,000 samples as the burn-in period, and retain every 50th sample from the remaining samples to approximate samples of the posterior distribution. This result serves as a reference solution for this tomographic problem. Figure \ref{fig:tomo_low_vel_mcmc} shows the mean and standard deviation maps obtained using MH-McMC. We find that the mean models obtained from BVI and MH-McMC show similar results, and the uncertainty maps from both methods exhibit similar loop-like higher uncertainty structures. However, the uncertainties from BVI are slightly lower than those from MH-MCMC, indicating that BVI still underestimates the true uncertainty to some extent. Nevertheless, since BVI yields results comparable to MH-MCMC (which is often assumed to provide the true solution), we conclude that BVI provides an approximately correct and, more importantly, fully analytic result. Furthermore, it significantly improves upon the results obtained using mean field ADVI, and is obtained with far fewer forward evaluations than the Monte Carlo method.

\begin{figure}
	\centering\includegraphics[width=\textwidth]{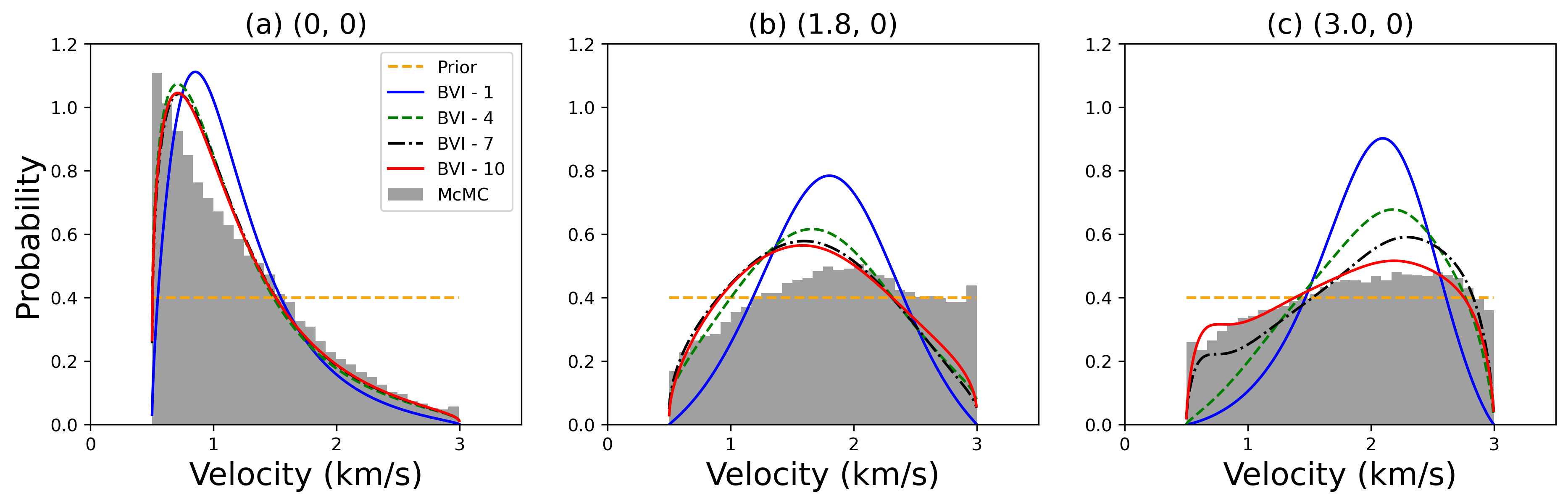}
	\caption{Marginal posterior distributions of velocity at three points located at (a) (0, 0) km, (b) (1.8, 0) km and (c) (3.0, 0) km, marked by three black crosses in Figures \ref{fig:tomo_low_vel_mean_std} and \ref{fig:tomo_low_vel_mcmc}. In each figure, the grey histogram shows the marginal distribution obtained using MH-McMC, and dashed yellow line shows the prior distribution. Blue, dashed green, dashed black, and red lines show marginal distributions obtained using BVI with 1 component (corresponding to mean field ADVI), 4, 7 and 10 components, respectively.}
	\label{fig:tomo_low_vel_marginal}
\end{figure}

In Figures \ref{fig:tomo_low_vel_marginal}a -- \ref{fig:tomo_low_vel_marginal}c, we compare the marginal distributions of three representative points at (0, 0) km, (1.8, 0) km and (3.0, 0) km. These locations are denoted by black crosses in Figures \ref{fig:tomo_low_vel_mean_std} and \ref{fig:tomo_low_vel_mcmc}. The first point lies within the low velocity anomaly, the second point is at the edge of the anomaly where the inner higher uncertainty loop is observed, and the last point is in the outer higher uncertainty loop. In each figure, the grey histogram shows the marginal distribution obtained using MH-McMC for reference, and dashed yellow line shows the Uniform prior pdf. For BVI, we can calculate the analytic marginal pdfs for these three points  without drawing any samples. Results using 1 BVI component (mean field ADVI), 4 components, 7 components, and 10 components are depicted by blue, dashed green, dashed black, and red lines, respectively. It is evident that mean field ADVI underestimates the posterior uncertainties, particularly in Figures \ref{fig:tomo_low_vel_marginal}b and \ref{fig:tomo_low_vel_marginal}c. However, as we add more components to the mixture, the marginal pdfs become increasingly similar to those obtained from MH-McMC, especially for the third point in Figure \ref{fig:tomo_low_vel_marginal}c, where the red line perfectly matches the grey histogram. In Figure \ref{fig:tomo_low_vel_marginal}b the results obtained using BVI and McMC are a little different. The reason might be that BVI components get stuck around a local mode, or that the Monte Carlo solution has not converged, since detecting convergence -- even with some well established methods such as the $\hat R$-statistic \cite{gelman1992inference} -- in problems of this dimensionality might be subjective. Therefore, it is difficult to be certain which one of these two results is better. Nevertheless, we still observe that each new component corrects some of the residual from the previous distributions in the ensemble, apparently boosting the accuracy of the current variational distribution (hence the name, ``boosting variational inference").

Figures \ref{fig:tomo_low_vel_mean_std} and \ref{fig:tomo_low_vel_marginal} show that the results achieve a reasonable approximation to the true posterior distribution using only 7 components. Unfortunately, in real problems we do not have access to the true posterior distribution, and running a McMC test for high-dimensional problems is often infeasible. Consequently, it becomes challenging to decide when to stop adding more components. A viable approach is to monitor the convergence of the KL-divergence: after each BVI iteration, we estimate KL[$q^t(\mathbf{m})||p(\mathbf{m}|\mathbf{d}_{obs})$] by drawing samples from the mixture distribution $q^t(\mathbf{m})$, and stop the BVI algorithm once KL[$q^t(\mathbf{m})||p(\mathbf{m}|\mathbf{d}_{obs})$] ceases to decrease. However, accurately estimating the KL-divergence for high-dimensional problems is hindered by the curse of dimensionality. In this example, we therefore compared statistical properties that can be estimated more stably such as the mean, standard deviation, and marginal pdf of the current mixture distribution with those obtained from previous iterations. If no significant changes are observed, we assume that BVI has converged and refrain from adding more components.

\subsection{Field Data Test}
\begin{figure}
	\centering\includegraphics[width=0.4\textwidth]{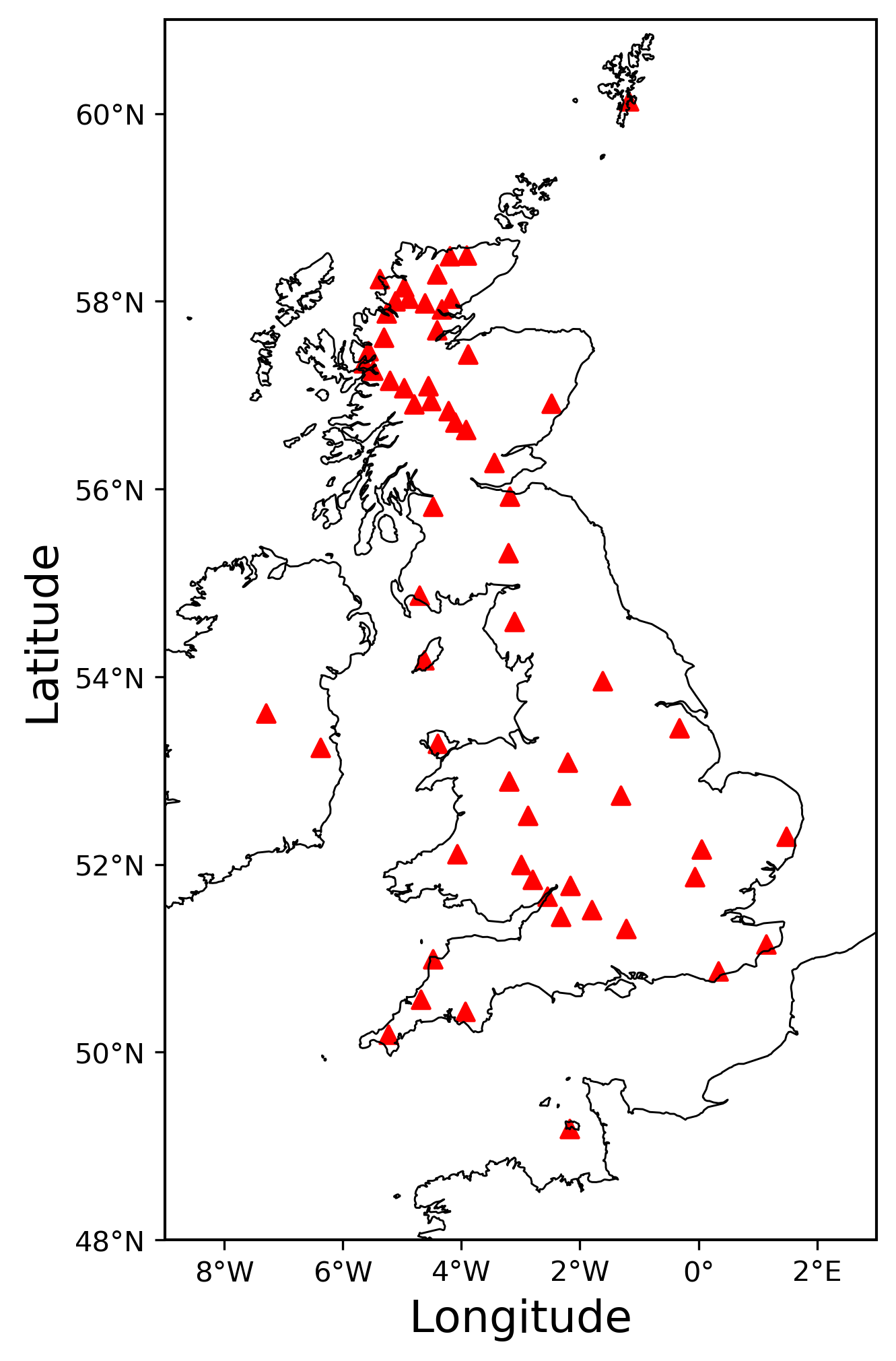}
	\caption{The location of 61 seismometers (red triangles) around the British Isles. The receiver stations are also treated as virtual sources for ambient noise interferometry to estimate inter-receiver first arrival travel times, which are used as the observed data in this test.}
	\label{fig:uk_receiver}
\end{figure}

In a more complicated field data example we applied BVI to Love wave tomography of the British Isles. The British Isles have been extensively studied and well understood using ambient noise tomography with different inversion methods, including linearised inversion \cite{nicolson2012seismic, nicolson2014rayleigh}, rj-McMC \cite{galetti2017transdimensional} and variational inference \cite{zhao2021bayesian, zhao2022interrogating}. Therefore, this is a suitable real-data test case to evaluate the performance of the proposed method and analyse the results by comparison. We use part of the dataset created by \citet{galetti2017transdimensional}: ambient noise data recorded by 61 seismometers located around the British Isles, as indicated by red triangles in Figure \ref{fig:uk_receiver}. The data were collected during three periods: 2001--2003, 2006--2007, and in 2010. The two horizontal components of the data were cross-correlated to compute Love waves between pairs of receiver stations. Subsequently, the first arrival travel times of group velocity were estimated at different periods ranging from 4 s to 15 s. Detailed information regarding the station network and data processing procedures can be found in \citet{galetti2017transdimensional}. For this test, we use the travel time measurements of Love waves at period 10 s.

We parametrise the target region in Figure \ref{fig:uk_receiver} into 37 $\times$ 40 regular grid cells with a spacing of 0.33$^\circ$ in both longitude and latitude directions. For each grid cell, we define a Uniform prior distribution ranging from 1.56 to 4.84 km/s: the average value of the Uniform distribution is obtained by measuring the average velocity across all valid ray paths by assuming a homogeneous medium, and the upper and lower bounds are chosen to exceed the range of velocities observed on the dispersion curves. The likelihood function is chosen to be a Gaussian distribution, and the travel time uncertainty for each inter-receiver path is estimated from the standard deviation of the estimated travel time of the corresponding station pair constructed by stacking randomly selected subsets of daily cross-correlations \cite{galetti2017transdimensional}.

Given this problem's higher dimensionality (1480) and non-linearity (due to higher noise and irregular data distribution) compared to the 2D synthetic test, BVI requires more components to converge to a reasonable approximation of the true posterior distribution. However, the greedy algorithm described in previous sections is time-consuming and does not fully use the computational power of modern compute clusters. To address this we propose an efficient implementation of BVI by running multiple independent runs in parallel, similar to McMC methods that often run independent chains in parallel. In this implementation, we start each independent BVI run from the second component, as the first component (corresponding to ADVI) has been shown to provide a stable (though inaccurate) result \cite{kucukelbir2017automatic, zhang2019seismic}. Each independent BVI run is initialised randomly and optimised separately, and after optimisation, the mixture distributions obtained from all runs are averaged to obtain the final approximation to the posterior distribution. This parallelisation approach allows BVI to take advantage of parallel computing capabilities while still providing analytic results.

We apply BVI and MH-McMC to this tomography problem for comparison. We run 4 independent BVI tests in parallel, and for each test we use 5 components. This results in a total of 20 Gaussian distributions used to approximate the first five Gaussian components that best fit the posterior distribution. Again, we use a diagonal Gaussian distribution as the mixture component. Each component is updated for 5000 iterations with 2 samples used at each iteration. The weight coefficients for the mixture components are calculated using equation \ref{eq:weight_1}. After optimisation, we average the distributions obtained from the 4 runs to obtain the final results. To obtain results using MH-McMC, we run 10 Markov chains in parallel. Each chain consists of 1.5 million samples, with the first 1 million samples discarded as burn-in. We discard a large number of samples in the hope that the remaining samples are reasonably well distributed according to the posterior distribution. After the burn-in period every 100th sample is retained to approximate an ensemble of posterior samples.

\begin{figure}
	\centering\includegraphics[width=\textwidth]{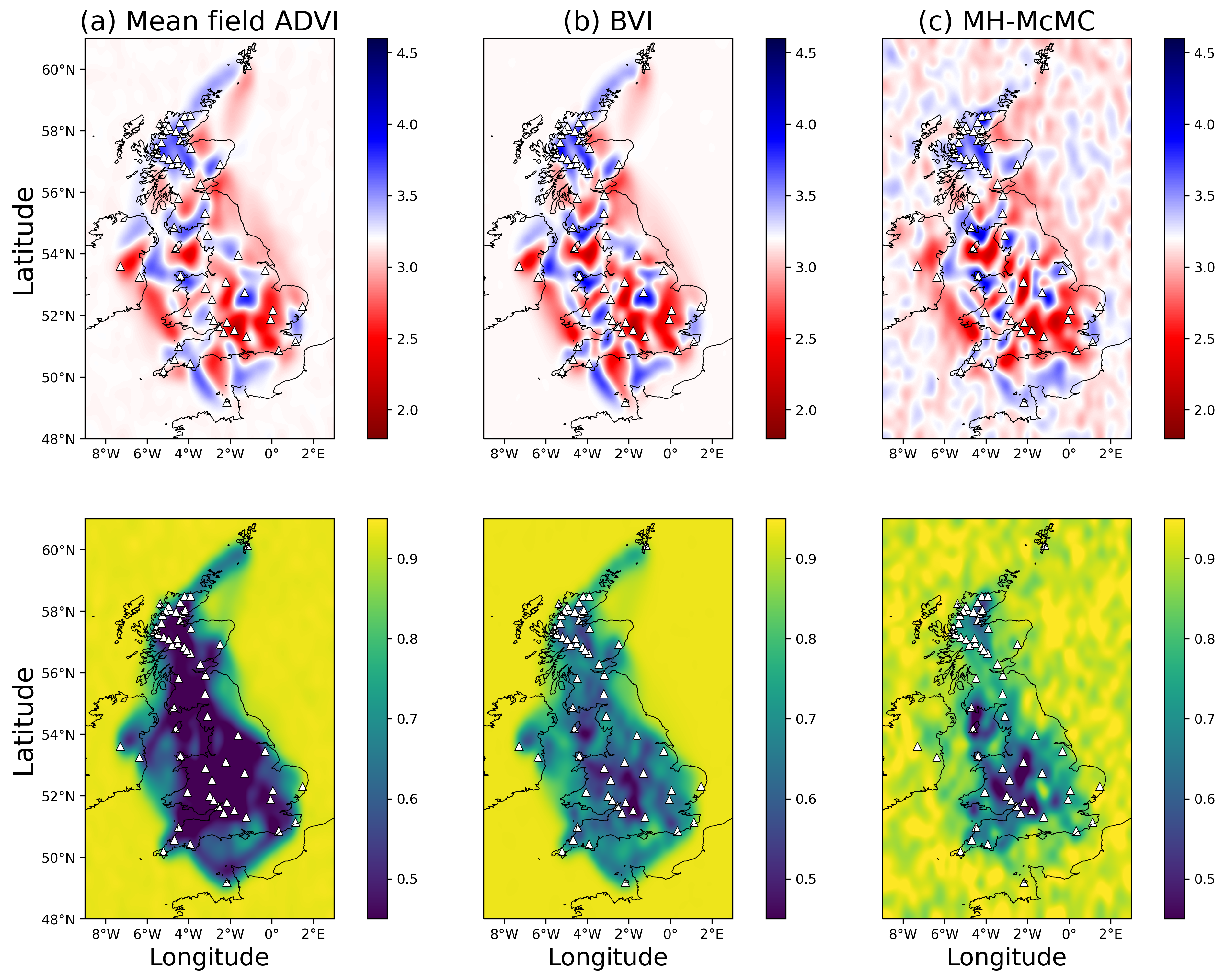}
	\caption{Mean (top row) and standard deviation (bottom row) maps of the Love wave tomography results of the British Isles using (a) mean field ADVI, (b) BVI, and (c) MH-McMC. White triangles show the locations of the receivers used in this example.}
	\label{fig:uk_mean_std}
\end{figure}

Figures \ref{fig:uk_mean_std}b and \ref{fig:uk_mean_std}c show the average velocity (top row) and standard deviation (bottom row) maps of the Love wave tomography results obtained using BVI and MH-McMC. We also display the results obtained using mean field ADVI, which corresponds to the first component obtained from BVI, as shown in Figure \ref{fig:uk_mean_std}a. The average velocity maps from the three methods exhibit similar features that are consistent with the known geology of the British Isles. For example, we observe a high velocity anomaly in the Scottish Highlands (6$^\circ$W -- 4$^\circ$W and 57$^\circ$N -- 59$^\circ$N), reflecting the crystalline metamorphic origin of the rocks in that region. A low velocity structure is observed in the area between 5$^\circ$W -- 3$^\circ$W and 53$^\circ$N -- 55$^\circ$N, which corresponds to the East Irish Sea sedimentary basins. Several low velocity anomalies are also observed around the Midland Platform in southern England (3$^\circ$W -- 1$^\circ$E and 50$^\circ$N -- 52$^\circ$N), corresponding to various sedimentary basins such as the Cheshire Basin, the Anglian-London Basin, and the Wessex Basin. 

The uncertainty models obtained from BVI and MH-McMC present similar patterns. For instance, lower uncertainties are observed in regions with densely placed receiver arrays such as across the Highlands and southern England. A higher uncertainty loop is observed around the East Irish Sea (4$^\circ$W and 54$^\circ$N) since a wide variety of different anomaly shapes and velocity values fit the observed travel time data, which is consistent with the findings from previous studies \cite{galetti2017transdimensional, zhao2021bayesian}.

The results obtained from BVI and MH-McMC are similar to those from other variational methods -- normalising flows and Stein variational gradient descent (SVGD) in \citet{zhao2021bayesian}. However, there are some small differences in the structures observed in Figures \ref{fig:uk_mean_std}b and \ref{fig:uk_mean_std}c compared to those obtained from rj-McMC in \citet{galetti2017transdimensional}, which can be attributed to the different parametrisations used in that work (Voronoi cells versus regular cells). In the rj-McMC study \cite{galetti2017transdimensional}, 16 chains and 3 million samples per chain were used to ensure convergence. In this test, 10 chains and 1.5 million samples were used for MH-McMC. The presence of some non-smooth structures in Figure \ref{fig:uk_mean_std}c compared to the smooth structures in the synthetic test (Figure \ref{fig:tomo_low_vel_mcmc}) suggests that the chains may not have fully converged even after 1.5 million samples, and that 10 chains might not be sufficient to explore all possible parameter subspaces that fit the data. In \citet{zhao2021bayesian}, full rank ADVI was also applied to this problem. However, both full rank ADVI in that work and mean field ADVI here, exhibit strong biases in the uncertainty results, with lower uncertainty than the McMC results observed everywhere inside the receiver array. In conclusion, since similar solutions have been obtained by multiple different methods, it can be assumed that BVI is capable of providing a reasonable estimate of the posterior distribution with an analytic expression, while also improving performance compared to mean field ADVI.


Table \ref{table:uk} compares the computational costs associated with several different methods, measured in terms of the required number of forward evaluations, since forward simulation is the most expensive part in each inversion. The computational costs of full rank ADVI, normalising flows and SVGD are obtained from \citet{zhao2021bayesian}, while the cost of rj-McMC is obtained from \citet{galetti2017transdimensional}. For BVI, four parallel tests with five components are run, each updated for 5000 iterations with two samples per iteration. However, since the first component (mean field ADVI) is very stable, it only needs to be trained once, resulting in a total of 170,000 forward evaluations for BVI and 10,000 for mean field ADVI. MH-McMC consists of 10 chains with 1.5 million samples each, resulting in a total of 15 million samples. It should be noted that the comparison depends on subjectively detecting the convergence of each method and may not reflect the minimum possible computational cost. \citet{zhao2021bayesian} showed that the same MH-McMC restricted to 2 million samples only provides a few of the main features in the mean velocity model and hardly provides any useful information in the standard deviation map. This decreases the likelihood that our subjective assessment of when the Monte Carlo method had converged led to the large number of samples attributed to the method above, and justifies that the number of samples used for MH-McMC may be reasonable. It is also true that significantly more efficient Monte Carlo methods may exist for this problem. Nevertheless, the significantly different numbers in Table \ref{table:uk} provide valuable insights into the amount of computation that we and other authors judged necessary to approach convergence for each method. Both mean field ADVI and full rank ADVI have the lowest computational costs, but they also provide biased results. Normalising flows are slightly more efficient than BVI, but they require a sophisticated design of flow structures and often rely on neural networks \cite{dinh2014nice, dinh2017density, kingma2016improved, papamakarios2017masked, durkan2019neural}, which can be challenging or even impossible to train for high-dimensional problems such as full waveform inversion. BVI has a simpler structure, and each component is optimised sequentially, making it more attractive for large scale datasets with higher dimensionality in real applications. SVGD is the most expensive variational method tested, but it still offers a significant reduction in cost compared to the two Monte Carlo methods. The huge numbers of samples used in the latter methods indicate a significant efficiency improvement offered by variational inference for solving large scale and high dimensional inverse problems. 

\begin{table}
	\caption{Number of forward evaluations required for different methods to provide the Love wave tomography results across the British Isles. The results for full rank ADVI, normalising flows and SVGD are from \citet{zhao2021bayesian}, while the result for rj-McMC is from \citet{galetti2017transdimensional}.}
	\centering
	\begin{tabular}{cc}
		\hline
		Method  & Forward Evaluations  \\
		\hline
		Mean field ADVI  & 10,000   \\
		Full rank ADVI  & 10,000   \\
		Normalizing Flows  & 100,000   \\
		BVI  & 170,000   \\
		SVGD  & 600,000   \\
		MH-McMC  & 15,000,000   \\
		RJ-McMC  & 48,000,000   \\
		\hline
	\end{tabular}
	\label{table:uk}
\end{table}

\section{Full Waveform Inversion}
\subsection{Bayesian FWI Implementation}
Seismic full waveform inversion (FWI) is a powerful technique to image subsurface structures using full waveform information in seismic data \cite{tarantola1984inversion, tromp2005seismic}. It is a highly non-linear and non-unique problem. Traditional linearised inversion methods can not reliably offer accurate solutions or effectively estimate the uncertainties in the inversion results. As a result, it is important to employ fully non-linear inversion methods for FWI. 

FWI problems typically have high dimensionality, and the forward modelling step, in which synthetic seismic waveforms are computed for a given velocity model, is usually expensive. To address these challenges, several efficient Monte Carlo methods have been applied to FWI \cite{qin2016underground, ray2016frequency, ely2018assessing, visser2019bayesian, guo2020bayesian, gebraad2020bayesian, kotsi2020time, zhao2021gradient, biswas2022transdimensional, de2023acoustic}. Alternatively, in recent years variational methods have also been introduced to address the computational challenges of Bayesian FWI \cite{zhang2021bayesianfwi, wang2023re, zhang20233, lomas20233d}. However, none of these methods provide an accurate and analytic approximation to the posterior probability distribution. In this section, we apply the BVI method to Bayesian FWI, to test its ability to provide analytic results efficiently.

\begin{figure}
	\centering\includegraphics[width=\textwidth]{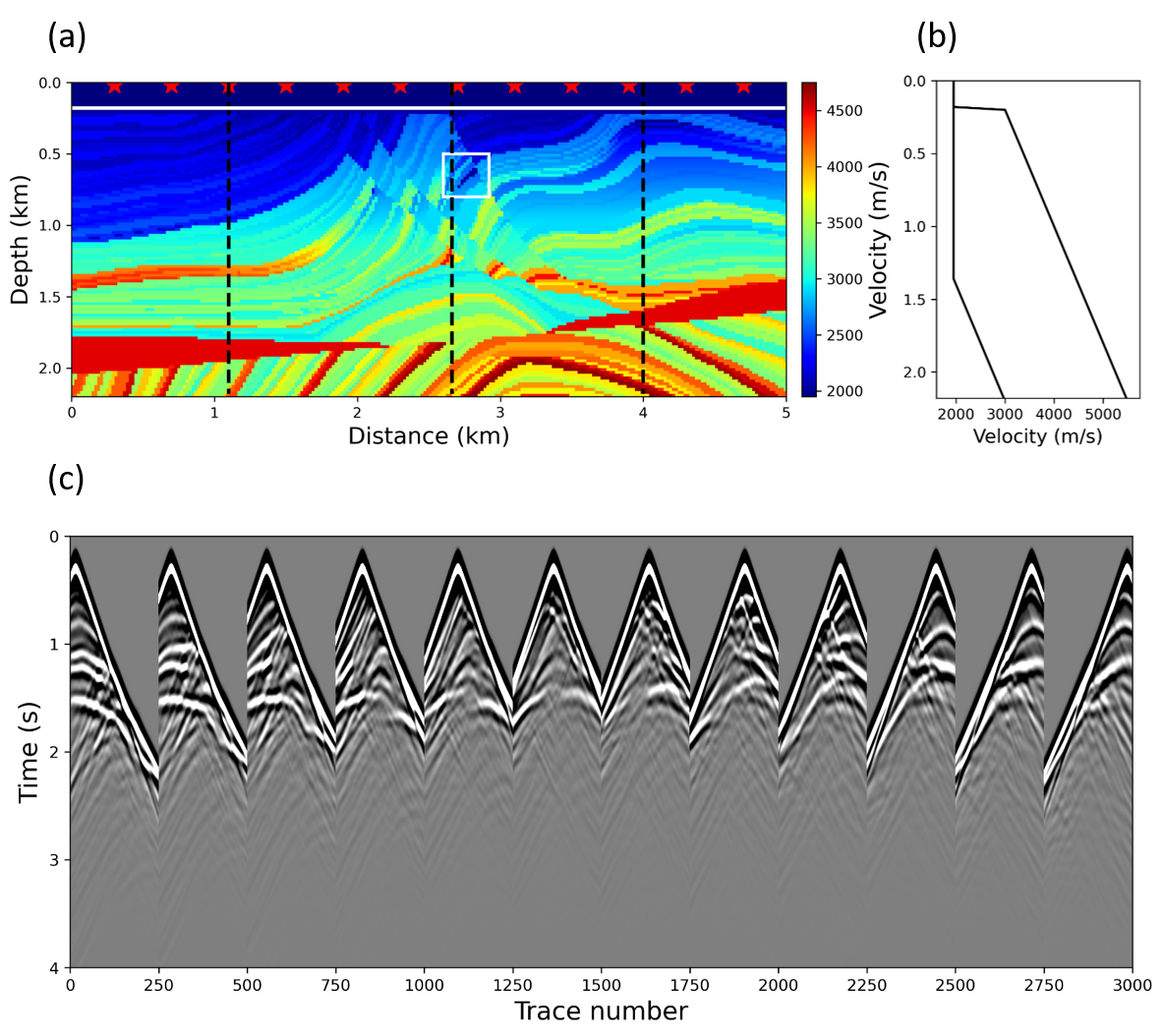}
	\caption{(a) The true Marmousi P wave velocity model with source locations indicated by red stars and receiver line marked by white line. Three dashed black lines display the locations of three well logs discussed in the main text. (b) Upper and lower bounds for the Uniform prior probability distribution for P wave velocity at each depth. (c) Twelve common shot gathers used as the observed data in this test.}
	\label{fig:fwi_110_250_vel_data_prior}
\end{figure}

We demonstrate the preceding BVI algorithm using a 2D acoustic FWI example. The true velocity model is a truncated Marmousi model \cite{martin2006marmousi2}, as shown in Figure \ref{fig:fwi_110_250_vel_data_prior}a. The density is set to be constant. The velocity field is discretized using 110 $\times$ 250 square grid cells with side length 20 m. Twelve sources are placed along the surface at 400 m intervals (shown by red stars in Figure \ref{fig:fwi_110_250_vel_data_prior}a), and 250 receivers are placed along the seabed at a depth of 200 m (white line in Figure \ref{fig:fwi_110_250_vel_data_prior}a). The observed waveform data are obtained by solving the 2D acoustic wave equation using the finite difference method, and the total simulation time is 4 s with a sample interval of 2 ms. The source is a Ricker wavelet with a dominant frequency of 5 Hz. Figure \ref{fig:fwi_110_250_vel_data_prior}c shows this waveform dataset.

For inversion, we use a Uniform prior distribution for the velocity model at each depth, with lower and upper bounds shown in Figure \ref{fig:fwi_110_250_vel_data_prior}b. Velocity in the water layer is fixed at the true value during inversion. Therefore, there are 25,000 free parameters to be inverted, corresponding to the subsurface velocity model. We use the finite difference method to solve the forward function, and the adjoint-state method to calculate the data-model gradient \cite{fichtner2006adjoint, plessix2006review}. The likelihood function is chosen to be a diagonal Gaussian with a constant data error of 0.05 around each datum.

In this test, we compare BVI with 3 different variational methods: mean field ADVI, Stein variational gradient descent (SVGD) and stochastic SVGD (sSVGD). Stochastic SVGD is an extension of SVGD that incorporates a noise term to enhance the efficiency and accuracy of SVGD for large-scale inference problems \cite{gallego2018stochastic}. It effectively converts the variational SVGD method to a Markov chain Monte Carlo method, showing that the divide between these methodological approaches can be bridged, and sSVGD has recently been applied to a 3D FWI problem \cite{zhang20233}. For mean field ADVI we use a diagonal Gaussian distribution to model the posterior distribution in the unconstrained space \cite{kucukelbir2017automatic}. A total of 50,000 hyper-parameters (means and variances in each cell) are updated for 10,000 iterations, and 5 samples per iteration are used. For SVGD, we randomly select 600 samples from the prior distribution and update them for 600 iterations. Once convergence is achieved, these samples are used to approximate statistics of the posterior distribution. For sSVGD, the algorithm starts with 24 random samples drawn from the prior distribution. These samples are then updated for 10,000 iterations, with the first 5,000 iterations discarded as the burn-in period. In this algorithm every sample value evaluated can be retained post burn-in, so all remaining samples are used to approximate the posterior distribution. For BVI, four parallel runs are performed, and each run contains six diagonal Gaussian distributions. This results in a total of 24 Gaussian components used to approximate the first six components that best fit the posterior distribution. Each component is updated for 5,000 iterations, and two samples per iteration are used.

\begin{figure}
	\centering\includegraphics[width=\textwidth]{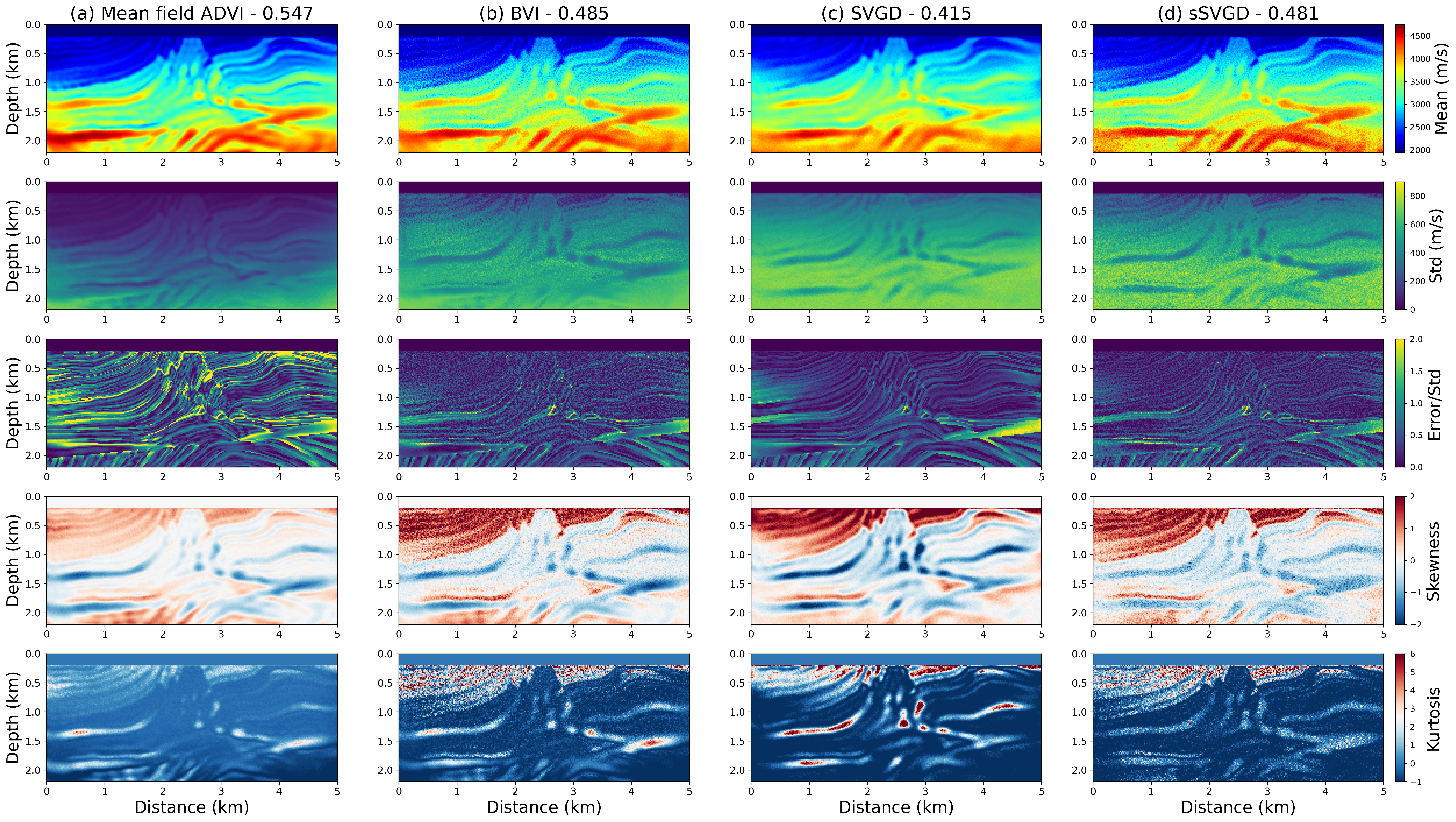}
	\caption{Mean, standard deviation, relative error, skewness, and kurtosis (from top to bottom row) of the posterior distribution for the 2D acoustic FWI test obtained using in column (a) mean field ADVI, (b) BVI, (c) SVGD and (d) sSVGD. The relative error is the absolute error between the mean and true models divided by the corresponding standard deviation. The number in the title of each column gives the Structural Similarity (SSIM) calculated between the mean and true models.}
	\label{fig:fwi_110_250_mean_std}
\end{figure}

Figures \ref{fig:fwi_110_250_mean_std}a -- \ref{fig:fwi_110_250_mean_std}d display the inversion results obtained using the aforementioned methods. The first two rows show the mean and standard deviation maps of the posterior distribution, while the third row displays the relative error, which is calculated by dividing the absolute error between the true and mean models by the standard deviation model. The mean velocity models from the 4 methods exhibit a similar pattern and generally resemble the true model. For example, within the white box in Figure \ref{fig:fwi_110_250_vel_data_prior}a, we observe a low velocity structure in the true and mean velocity models. However, all four mean velocity maps fail to capture some of the fine-scale structures present in the true model. This can be attributed to the low dominant frequency used in this test (5 Hz). Among the four methods, the mean velocities obtained using mean field ADVI and SVGD appear smoother compared to those obtained using BVI and sSVGD. This observation is consistent with the results obtained in the previous example of 2D synthetic travel time tomography, where the posterior distribution obtained using MH-McMC (Figure \ref{fig:tomo_low_vel_mcmc}) is smoother than that obtained using BVI (Figure \ref{fig:tomo_low_vel_mean_std}). In the case of BVI, since we use a diagonal Gaussian distribution as the component distribution, the mean estimate and uncertainty of each model parameter is updated independently. Every new component is initialised randomly to enhance the current posterior pdf by boosting it on either the lower or higher velocity side of the current mean estimate, and is optimised to approximate the posterior distribution within a local region in the parameter space, introducing a degree of variation between iterations. Hence, the results obtained from BVI exhibit less smoothness, despite the fact that results obtained from its first component (ADVI) are smooth. In sSVGD, the introduction of a noise term during each iteration perturbs the samples, leading to increased randomness \cite{zhang20233}. The result may therefore become smoother as a larger number of samples and iterations are used. 

We also calculate the Structure Similarity Index Model (SSIM) between each mean and the true velocity models, and display the calculated SSIM values along the top of Figures \ref{fig:fwi_110_250_mean_std}a -- \ref{fig:fwi_110_250_mean_std}d. SSIM is a common measure to quantify the similarity between two images, and higher values indicate higher similarity \cite{wang2004image, levy2022using}. Mean field ADVI provides a higher SSIM value, which means that the method is able to provide an accurate estimate of the mean model. However, it tends to underestimate the posterior uncertainties (see below). Among the other three methods, BVI and sSVGD provide similar and higher SSIM values compared to SVGD, potentially meaning that mean models estimated from these two methods are more accurate.

The standard deviation obtained using mean field ADVI significantly differs from the other three results and tends to be underestimated. Moreover, a majority of the relative errors are larger than 3, indicating inaccuracy of the results. However, the uncertainty map still exhibits similar geometrical structures compared to the mean and true velocity models. Therefore, we consider ADVI to be an efficient method that provides a fairly accurate mean model but biased uncertainties due to its restrictive theoretical assumptions \cite{zhang2019seismic, zhao2021bayesian}. Similarly to the mean velocity results, SVGD yields a smoother standard deviation map compared to BVI and sSVGD. In other aspects, the results obtained using these three methods are similar, with errors mainly distributed within two standard deviations of the mean. For example, we observe lower uncertainties and higher relative errors at locations with higher velocity anomalies (such as the higher velocity layer at a depth of 1.3 km depth and a distance between 0 -- 2 km). Additionally, higher uncertainties are observed at layer boundaries, which is consistent with our observations in the two travel time tomography examples. These correspond to uncertainty loops found in previous studies \cite{galetti2015uncertainty}, especially in the shallower subsurface where data exhibits higher sensitivity. However, the uncertainty values obtained using BVI are slightly smaller compared to the other two methods. We attribute this to two main factors. First, the use of a diagonal Gaussian distribution in BVI tends to underestimate the uncertainty information compared to a Gaussian distribution with a full covariance matrix \cite{kucukelbir2017automatic}. This underestimation of posterior uncertainties is also evident in Figures \ref{fig:tomo_low_vel_mean_std} and \ref{fig:tomo_low_vel_mcmc}. On the other hand, SVGD and sSVGD employ a repulsive force between different samples \cite{liu2016stein, gallego2018stochastic}: this pushes samples away from each other such that they can explore different parameter subspaces (while still approximating the posterior pdf with sample density). In cases where samples are sparsely distributed within the parameter space, as is the case in this test with 600 samples for SVGD and 24 samples per iteration for sSVGD, the repulsive force might push samples towards the corners of parameter hyperspace to maximise the objective function. This may also contribute to higher uncertainties in Figures \ref{fig:fwi_110_250_mean_std}c and \ref{fig:fwi_110_250_mean_std}d. A similar phenomenon was observed in Love wave tomography using SVGD \cite{zhao2021bayesian}. Given the absence of a true solution to this Bayesian FWI problem, it is challenging to determine which method provides a more accurate result. Nevertheless, obtaining similar results using three methods based on two different theoretical frameworks lends credibility to these findings.

In the 4th and 5th rows of Figure \ref{fig:fwi_110_250_mean_std} we show the skewness and kurtosis maps. These two statistics measure the third and fourth moment of the posterior distribution, where the skewness indicates the shift towards the left (negative skewness) or the right (positive skewness) of a distribution with respect to a normal distribution, and the kurtosis is a measure of the `bulkiness' (lower values) or `pointiness' (higher values) of a distribution \cite{scheiter2022upscaling}. We observe that the results from BVI, SVGD and sSVGD exhibit higher skewness and kurtosis values at the near surface and are similar to each other compared to those from mean field ADVI, which proves that these three methods capture (possibly correct) non-Gaussian structure of the posterior distribution.

\begin{figure}
	\centering\includegraphics[width=0.99\textwidth]{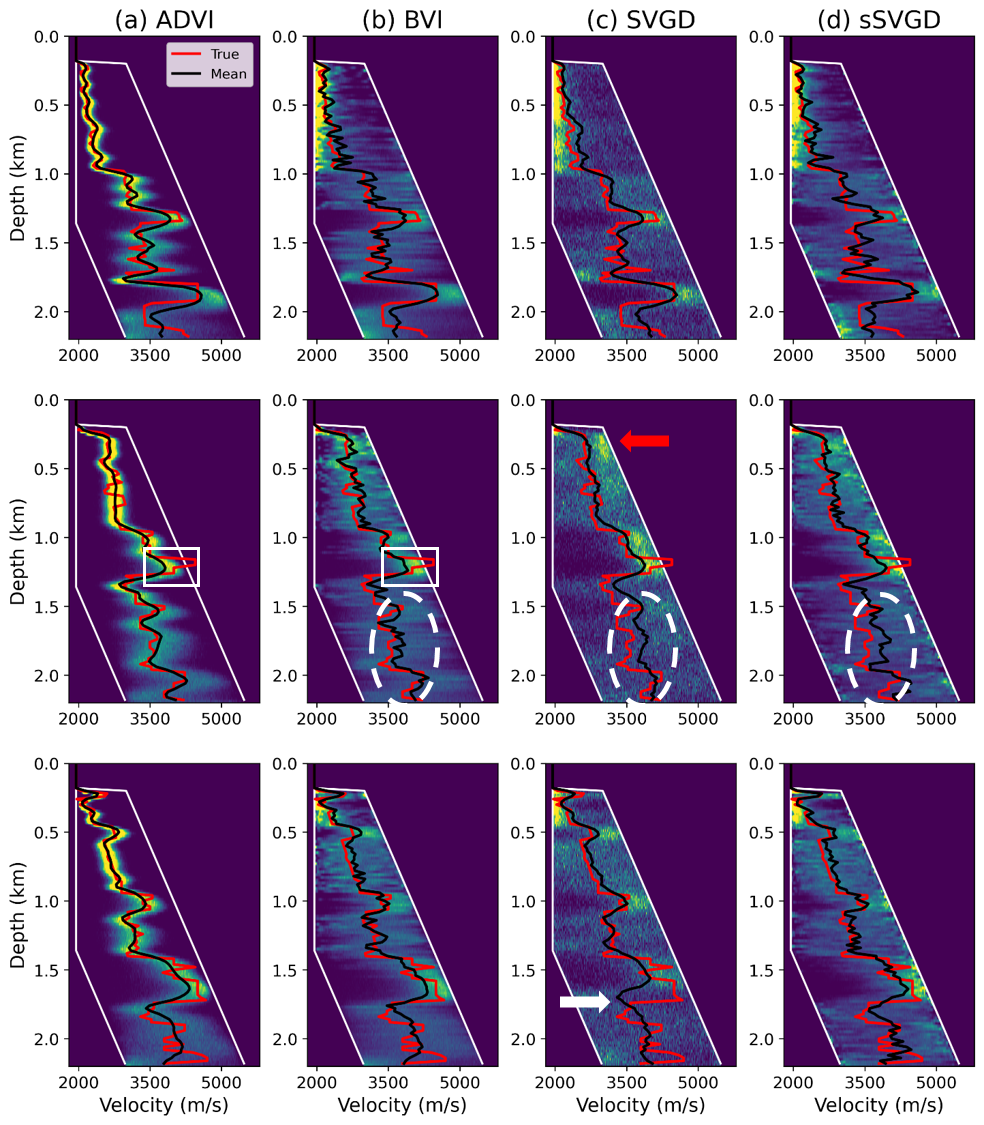}
	\caption{Marginal posterior distributions along vertical profiles at three locations (represented by black dashed lines in Figure \ref{fig:fwi_110_250_vel_data_prior}a) obtained using (a) mean field ADVI, (b) BVI, (c) SVGD and (d) sSVGD, respectively. Each row displays the marginal distribution along one profile. In each figure, two white lines show the prior bounds at each depth, the black line shows the mean velocity model, and the red line shows the true velocity model.}
	\label{fig:fwi_110_250_marginals}
\end{figure}

For better comparison, Figures \ref{fig:fwi_110_250_marginals}a -- \ref{fig:fwi_110_250_marginals}d display the marginal pdfs obtained using ADVI, BVI, SVGD and sSVGD, respectively, along three vertical profiles marked by dashed black lines in Figure \ref{fig:fwi_110_250_vel_data_prior}a. Each row shows the marginal distributions along one profile using the four methods. Red lines show the true velocity profiles and black lines show the mean velocity profiles obtained using each method. Similarly to the mean and standard deviation maps in Figure \ref{fig:fwi_110_250_mean_std}, ADVI provides accurate mean velocity profiles but underestimates posterior uncertainties, as evidenced by the narrower marginal pdfs compared to the other three methods. As discussed in the Methodology section, BVI boosts the results obtained from ADVI by using multiple Gaussian components to approximate the posterior distribution. This effect can be observed when comparing Figures \ref{fig:fwi_110_250_marginals}a and \ref{fig:fwi_110_250_marginals}b: BVI explores the parameter space that was not adequately represented by ADVI, resulting in wider (and potentially more accurate) marginal distributions. This is particularly noticeable at a depth of 1.2 km within the two white rectangular boxes in the second row in Figures \ref{fig:fwi_110_250_marginals}a and \ref{fig:fwi_110_250_marginals}b, where the true velocity value exceeds the prior upper bound (deliberately, to check performance in anomalous cases in which prior distributions are mis-specified). The posterior pdf obtained using BVI is concentrated closer to the upper bound of the prior distribution compared to ADVI. The marginal pdfs obtained using BVI and sSVGD are highly similar and slightly different from those obtained using SVGD. The results from SVGD are sparser (due to limited number of samples) and smoother. In the shallower part of the second row of Figure \ref{fig:fwi_110_250_marginals}c (indicated by a red arrow), the higher probability region of the posterior pdf from SVGD is located close to the prior bound and deviates from the true value. This might be caused by either the limited number of samples or the relatively large step size used in SVGD, which pushes samples towards the boundary of parameter space by the repulsive force. At a depth of 1.7 km in the third row of Figure \ref{fig:fwi_110_250_marginals}c (indicated by a white arrow), the mean velocity value deviates from the true value since SVGD fails to provide a sufficiently high resolution result to recover this high velocity anomaly compared to BVI and sSVGD. Additionally, as indicated by three dashed white boxes in the second row, the posterior distributions from SVGD and sSVGD cover a larger parameter space than that from BVI, especially around the high velocity region. Consequently, we observe higher standard deviation values in Figures \ref{fig:fwi_110_250_mean_std}c and \ref{fig:fwi_110_250_mean_std}d compared to Figure \ref{fig:fwi_110_250_mean_std}b. However, in this region, the mean velocity model obtained using BVI is more similar to the true model, which might indicate higher accuracy compared to both SVGD and sSVGD. This demonstrates that higher uncertainties provided by SVGD and sSVGD might be less convincing due to effects of the repulsive force, as previously discussed and observed in \citet{zhao2021bayesian}.

Finally, we compare the computational cost of the four methods in Table \ref{table:fwi}. In FWI, the forward simulation and data-model gradient calculation are much more expensive compared to those in travel time tomography. Therefore, the number of gradient evaluations provides a fair comparison. For mean field ADVI the model is updated for 10,000 iterations using 5 samples per iteration, resulting in 50,000 evaluations. In the case of BVI, we run 4 parallel tests, each containing 6 Gaussian components. However, we do not need to optimise the first component 4 times, thus a total of 21 Gaussian distributions are used. For each component, we use 5000 iterations and 2 samples per iteration. Therefore, BVI requires 210,000 gradient simulations. It is worth noting that the number of simulations used to optimise each component for BVI is smaller than that for ADVI, even though they have the same hyper-parameters (mean and standard deviation for a diagonal Gaussian distribution). This is because in BVI we do not require full convergence of each component. As long as new components fill some of the gap (residual) between the current mixture distribution and the true posterior distribution, this improves the current result. By adding more components, BVI gradually improves the posterior approximation. sSVGD and SVGD require 240,000 and 360,000 gradient evaluations, respectively. Overall, ADVI is the cheapest method, but it produces biased results. BVI requires more computations to improve the biased results from ADVI, and is slightly more efficient than sSVGD. More importantly, BVI provides an analytic solution to the posterior distribution, while sSVGD only provides posterior samples. SVGD is the most expensive method and only provides 600 samples, which is far from sufficient to approximate such a high dimensional (25,000) space in this test.

\begin{table}
	\caption{Number of forward and gradient evaluations for mean field ADVI, BVI, SVGD and sSVGD applied to the 2D FWI test. The values represent an indication of the computational cost of each method, as the evaluation of data-model gradients is the most computationally expensive part of this test.}
	\centering
	\begin{tabular}{cc}
		\hline
		Method  & Number of Gradient Evaluations \\
		\hline 
		ADVI  & 50,000   \\
		BVI  & 210,000\\
		SVGD  & 360,000   \\
		sSVGD  & 240,000   \\
		\hline
	\end{tabular}
	\label{table:fwi}
\end{table}


\subsection{Interrogating FWI Results}
\begin{figure}
	\centering\includegraphics[width=0.999\textwidth]{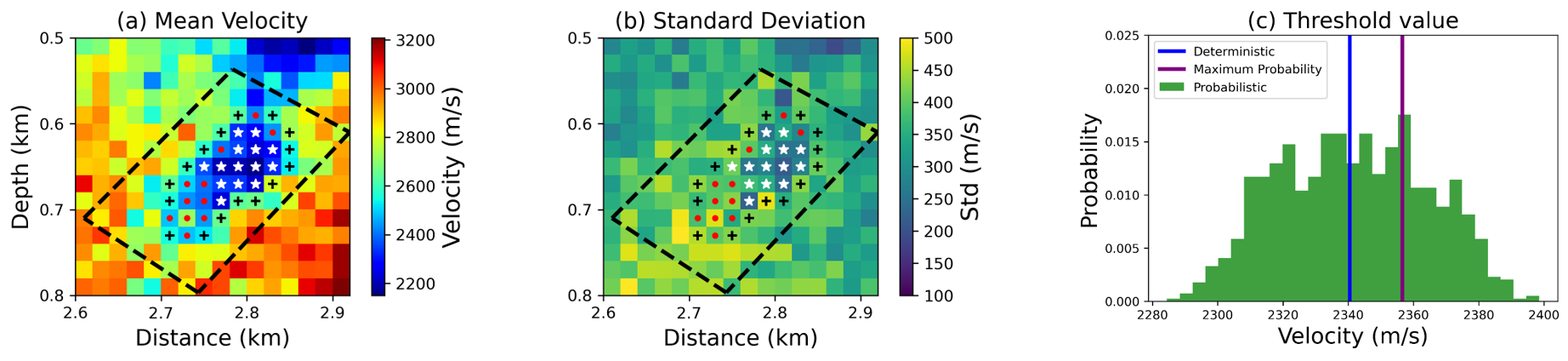}
	\caption{(a) Mean and (b) standard deviation maps of the posterior distribution obtained using BVI within the white box in Figure \ref{fig:fwi_110_250_vel_data_prior}a. Black dashed box shows the mask inside which we calculate the area of the largest continuous low velocity body, which serves as the target function. White stars and black crosses denote cells that are most likely to be inside and outside the reservoir, respectively. Red dots denote cells located on or near the reservoir boundaries, where uncertainty remains regarding their classification as low or high velocities. (c) Threshold values to discriminate low and high velocities. Green histogram shows the probabilistic threshold value established in the main text. Blue line shows the deterministic threshold value obtained using the white stars and black crosses only, and purple line shows the maximum probability threshold value from the green histogram.}
	\label{fig:fwi_interro_mean_std_threshold}
\end{figure}

We demonstrate the advantages of the BVI solution for interrogation problems using the theory in section 2.4, by using the FWI results to answer a specific question: \textit{what is the size of the low velocity volume within the white box in Figure \ref{fig:fwi_110_250_vel_data_prior}a?} Such inquiries are common in the geoscience community and are used, for example, to estimate the volume of a sedimentary basin or the size of an ore body or a reservoir resource assessment or $CO_2$ storage \cite{fletcher1996estimation, burshtein2006statistical, romdhane2014co2, zhao2022interrogating, zhang2021interrogation}. We therefore denote the low velocity volume as a reservoir hereafter. Figures \ref{fig:fwi_interro_mean_std_threshold}a and \ref{fig:fwi_interro_mean_std_threshold}b show the posterior mean and standard deviation maps inside the white box, obtained using BVI.

Previously, volume-related questions were answered using interrogation theory with a deterministic target function in \citet{zhao2022interrogating} and \citet{zhang2021interrogation}. Here we provide a brief overview of the procedure. We first introduce a mask to restrict the region used to calculate the low velocity anomalies, as illustrated by the dashed black box in Figures \ref{fig:fwi_interro_mean_std_threshold}a and \ref{fig:fwi_interro_mean_std_threshold}b. Other low velocity bodies outside of this mask are assumed to be unrelated to the anomaly of interest and are ignored during the interrogation process. Considering a reservoir should be a continuous geological body in space, we define the target function to be the area of the largest continuous low velocity body inside the mask. To evaluate this function, we need to distinguish between low velocity and high velocity cells, which can be accomplished by introducing a threshold value: cells with velocity values below the threshold are classified as low velocity, others are classified as not low velocity.

\begin{figure}
	\centering\includegraphics[width=0.999\textwidth]{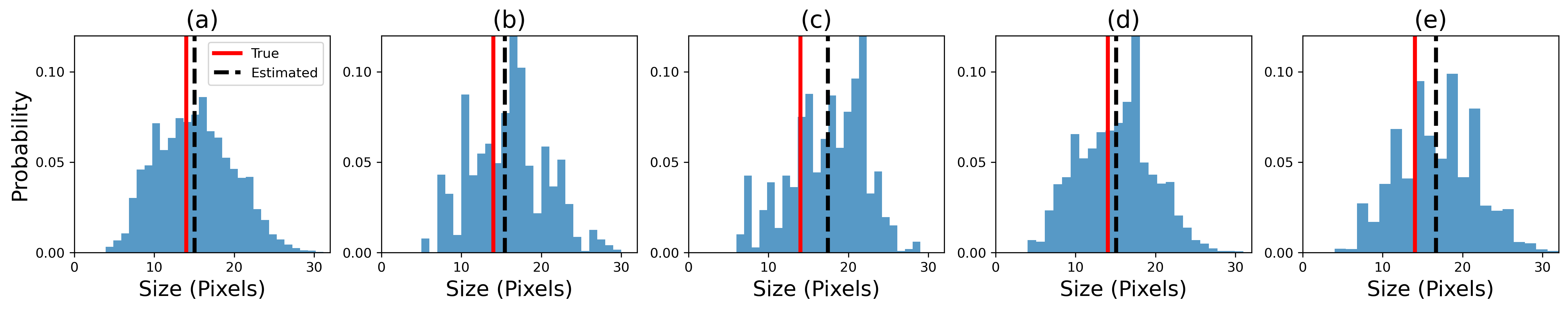}
	\caption{Posterior distributions of the target function by interrogation with probabilistic target function obtained using (a) full BVI inversion results, (b) 24 representative samples from BVI components, and (c) 40 random samples from the full BVI inversion results. Panels (d) and (e) show the posterior target functions obtained using deterministic target functions whose threshold values are represented by the blue and purple lines in Figure \ref{fig:fwi_interro_mean_std_threshold}c. In each figure, the red line denotes the true answer to this question, and black dashed line denotes the optimal answer obtained using interrogation theory.}
	\label{fig:fwi_interro_optimal_answer}
\end{figure}

We use the same data-driven method as \citet{zhao2022interrogating} to calculate the threshold value with minimal bias. First, we select a set of points from the inversion results that are most likely to belong to the low velocity reservoir since they have low mean velocity values and low standard deviation values (indicated by white stars in Figures \ref{fig:fwi_interro_mean_std_threshold}a and \ref{fig:fwi_interro_mean_std_threshold}b), and another set of points likely to be outside of the reservoir (represented by black crosses in Figures \ref{fig:fwi_interro_mean_std_threshold}a and \ref{fig:fwi_interro_mean_std_threshold}b). Then we calculate the average marginal cumulative density function (cdf) of the white stars being classified as inside the reservoir, accumulating as the velocity threshold increases, and of the black crosses being classified as outside of the reservoir, accumulating as the velocity threshold decreases. The intersection point of these cdfs is the threshold value that discriminates low from high velocities with minimal bias according to the prior information provided by the locations of white stars and black crosses. The corresponding threshold value is illustrated by the blue line in Figure \ref{fig:fwi_interro_mean_std_threshold}c. Given this value we can classify each cell as either a low or high velocity cell, find the largest continuous low velocity body inside the mask, and calculate its size which is the target function value. Figure \ref{fig:fwi_interro_optimal_answer}d shows the posterior distribution of the target function (reservoir size) obtained using this threshold value. According to equation \ref{eq:optimal_answer}, the optimal (minimum bias) answer is the mean of the target function values from all posterior samples, and is denoted by dashed black line in Figure \ref{fig:fwi_interro_optimal_answer}d. For comparison, the true size is denoted by a red line in Figure \ref{fig:fwi_interro_optimal_answer}d. 

The above method calculates the threshold value and the target function deterministically. As stated in section 2.4, this does not consider the uncertainty introduced by human bias, which may result in different sets of low and high velocity cells being selected by different experts, thus different threshold values and different target functions, potentially biasing reservoir size estimates. We therefore also apply interrogation with a probabilistic target function, which is defined by a probabilistic threshold value in this example. We implement this by randomly selecting a subset of the grid cells from each of the low and high velocity cells in Figures \ref{fig:fwi_interro_mean_std_threshold}a and \ref{fig:fwi_interro_mean_std_threshold}b. This random selection simulates possible variation in the selection by different experts. We also consider other cells situated on the boundaries of the low velocity body, as indicated by red dots in Figures \ref{fig:fwi_interro_mean_std_threshold}a and \ref{fig:fwi_interro_mean_std_threshold}b which in fact contain valuable information about reservoir shape and velocity values \cite{galetti2015uncertainty}, and incorporate the information provided by these cells to calculate the probabilistic threshold value. To do that, we randomly select a subset of cells marked by those red dots, and assign cells that are directly connected to the cells marked by the white stars as low velocity cells (inside the reservoir) and the remaining cells as high velocity cells (outside the reservoir). This can be interpreted as a misclassification of low and high velocity cells at the boundaries of the reservoir, again simulating possible human bias and subjective choice. We use these randomly selected cells to calculate the threshold value. The above procedure is repeated 1000 times, resulting in a probability distribution over the threshold value represented by the green histogram in Figure \ref{fig:fwi_interro_mean_std_threshold}c.



We perform interrogation using the green histogram in Figure \ref{fig:fwi_interro_mean_std_threshold}c as the stochastic threshold value, which then defines the probabilistic target function. For each posterior model sample (velocity model obtained from BVI), we draw 100 random threshold values from the green histogram and calculate the size of the largest continuous low velocity body corresponding to each threshold value. The resulting distribution of 100 reservoir sizes values incorporates the uncertainty in the target function, so we repeat this process for each posterior model sample. Figure \ref{fig:fwi_interro_optimal_answer}a shows the distribution of the target function values, and the optimal answer calculated using equation \ref{eq:probabilistic_interrogation} is denoted by the dashed black line. This represents the interrogation results (with the probabilistic target function) obtained using the full posterior distribution from BVI. We also construct a solution using only the representative samples obtained from the means of the BVI components to perform interrogation, and the posterior target function is displayed in Figure \ref{fig:fwi_interro_optimal_answer}b. The optimal answer is calculated using equation \ref{eq:probabilistic_optimal_answer_bvi} (black dashed line in Figure \ref{fig:fwi_interro_optimal_answer}b). Since we only use the mean vectors of the Gaussian components to obtain those representative samples, without considering the corresponding covariance matrices, the uncertainty of the posterior target function might be underestimated. Nevertheless, this still provides an accurate optimal answer while significantly reducing the number of target function evaluations. Additionally, we randomly choose 40 posterior samples from the full BVI inversion result and conduct probabilistic interrogation on these: the resulting posterior histogram is displayed in Figure \ref{fig:fwi_interro_optimal_answer}c. In comparison to Figure \ref{fig:fwi_interro_optimal_answer}b, the optimal answer obtained from this set of 40 samples is notably inaccurate, whereas of the order of ten representative samples from BVI provide an almost equally accurate interrogation answer as that from the full posterior solution. This proves the value of these representative samples.

To simulate the bias that may be introduced by using a deterministic target function, for example one defined by a single expert, we choose the maximum probability value from the green histogram in Figure \ref{fig:fwi_interro_mean_std_threshold}c as the threshold value (denoted by purple line in Figure \ref{fig:fwi_interro_mean_std_threshold}c). This value falls within the high probability region and can be treated as a reasonable threshold value obtained from one expert. We perform interrogation using this single threshold value, and the result is displayed in Figure \ref{fig:fwi_interro_optimal_answer}e. The optimal answer (black dashed line) shows a larger error and deviates more from the true answer than any estimate other than that from 40 random samples of the model posterior distribution in Figure \ref{fig:fwi_interro_optimal_answer}c.

Overall, the comparison of the five histograms in Figure \ref{fig:fwi_interro_mean_std_threshold} reveals that interrogation using deterministic target functions may yield biased results due to the subjective nature of human interpretation. This bias can be mitigated by using probabilistic target functions. Note that the optimal answer using the deterministic target function in Figure \ref{fig:fwi_interro_optimal_answer}d also provides an accurate result, and the posterior target function is similar to that in Figure \ref{fig:fwi_interro_optimal_answer}a. However, we usually do not know the true answer to our question in real problems, and therefore have no means to prioritise the answer from one interpretation over any other. Probabilistic interrogation considers the subjectivity from different experts, and provides a more convincing answer. The optimal answer obtained using representative samples from BVI components is accurate, which proves that these samples capture a key portion of the uncertainty information in the inversion results. In contrast, a similar number of randomly selected posterior samples fails to adequately represent this uncertainty. Therefore, subsequent uncertainty analysis tasks, especially those that are computationally intractable to perform for thousands of posterior samples (e.g., reservoir simulation), could be more efficiently carried out using the representative samples obtained from BVI.

\section{Discussion}

In our synthetic travel time tomography example, we provide a reliable criterion for assessing the algorithm's convergence. Ideally, one could assess the convergence of BVI at each iteration by evaluating the KL-divergence between the true posterior pdf and the current mixture distribution. However, accurate estimation of the KL-divergence for high-dimensional problems is very expensive. Since we obtain a set of representative samples from the mean vectors of the Gaussian components, we can use these samples to estimate the KL-divergence approximately, so as to provide a more reasonable way to detect convergence. This is similar to the idea used for probabilistic interrogation in equations \ref{eq:optimal_answer_bvi} and \ref{eq:probabilistic_optimal_answer_bvi}. 


As stated by the No Free Lunch theorem \cite{wolpert1997no}, no method is better than any other method when averaged across all problems, so there is no possibility to find a `best' method in general. However, for a particular class of problems it is possible to find better or worse suited algorithms from different points of view. In all of our examples, ADVI yields biased uncertainty results, but provides an accurate mean velocity map and is the most computationally efficient method. The first component of BVI can be regarded as equivalent to ADVI, and so establishes an estimate of the mean. BVI then introduces additional components to better approximate uncertainty in the true posterior distribution, trading off with a higher computational cost. In addition, BVI is parametric which allows an arbitrary number of samples to be generated essentially for free post optimisation, whereas this is far more expensive for sampling based methods (SVGD and sSVGD).

One possible improvement for BVI is to use Gaussian components with a full covariance matrix, but this can be computationally cumbersome for high-dimensional problems such as FWI, as it requires $D(D+1)/2$ hyper-parameters for a D-dimensional covariance matrix. Considering that only a few pairs of variables may exhibit significant posterior correlations (e.g., neighbouring cells), a feasible approach is to approximate the full covariance matrix using a low-rank plus diagonal approach \cite{miller2017variational}. 


Normalising flows are another variational method which can effectively model posterior correlations between different parameters \cite{dinh2014nice, dinh2017density, kingma2016improved, papamakarios2017masked}. It has been demonstrated that normalising flows outperform ADVI \cite{zhao2021bayesian}, making them a promising choice for improving BVI. By using probability distributions modelled by normalizing flows as the component distributions in BVI, we might capture posterior correlations and create model that enhances the capabilities of existing normalising flows while reducing the complexity for designing flows structures, albeit at the expense of greedy optimisation \cite{giaquinto2020gradient}.

Gaussian processes (GP) form another class of methods that use Gaussian distributions to approximate the probability distribution of model parameters. GP is a form of stochastic process, and can be regarded as a way to define a Gaussian distribution over functions (for example, to define Gaussian distributions for velocity values at every subsurface location). It is commonly used as a non-parametric regression method that predicts model parameters and the corresponding uncertainties within a continuous region. \citet{ray2019bayesian}, \citet{ray2021bayesian} and \citet{blatter2021two} used GP together with a trans-dimensional McMC sampling scheme to perform inversion. In those works GP was used as a regression method to build a finely discretized or even spatially continuous (infinite-dimensional) model vector $\mathbf{m}$, which can be viewed as a random sample from an infinite-dimensional multivariate Gaussian distribution, given parameter values at some known locations. The obtained model was used to calculate the synthetic data to further update the GP. \citet{valentine2020gaussian, valentine2020gaussian2} used GP to solve linear (or weakly non-linear) inverse problems. The inversion result can be expressed as a GP which represents the posterior distribution in function space. Due to the nature of GP, these works assume a Gaussian prior distribution for the model parameter at each location and a linear forward function \cite[as in][]{valentine2020gaussian}. Such assumptions are not necessary for BVI as described in this paper.

Making use of the analyticity of BVI results can be challenging, but we have developed an implicit approach to address this issue. Our approach involves selecting one representative sample from each BVI component: leveraging the fact that a parametric and symmetric Gaussian distribution is used as the component distribution. We simply adopt the mean vector as a representative sample, allowing us to obtain tens of samples directly that partially represent the posterior distribution for uncertainty analysis. Considering that we also obtain a diagonal covariance matrix for each component, it is easily possible to incorporate the information from the covariance matrix into these representative samples (for example, by selecting a number of component samples that is proportional to the weight of that component and combining all such samples). This would capture more detail from the posterior distribution and improve the effectiveness of uncertainty analysis.

In our interrogation example, we show that the optimal answer to an interrogation problem obtained using the representative samples is accurate and comparable to that obtained using full inversion results. This is particular attractive when implementing probabilistic interrogation as proposed in this paper, or when the evaluation of the target function is computationally expensive. For example, if our goal is to estimate $CO_2$ saturation of a reservoir using FWI results, the target function might involve reservoir simulation or (non-linear) rock physics inversion to convert seismic velocity values into $CO_2$ saturation. Calculating the target function for thousands of posterior samples could then be prohibitively expensive. In such cases, we can simply use the representative samples obtained from BVI components for analysis. Moreover, storing a large set of posterior samples on disk and loading them into memory can be extremely demanding, especially for 3D FWI problems \cite{zhang20233, lomas20233d}, to which the use of representative samples or the fully analytic and parametric posterior expression from BVI results provides a practical solution, as demonstrated in \citet{scheiter2022upscaling}. Finally, it is important to note that obtaining these representative samples would be challenging without the analytic expression of the posterior distribution provided by BVI, which provides these samples directly.

\section{Conclusion}
We have presented boosting variational inference (BVI) as a powerful variational method for solving fully non-linear Bayesian geophysical inverse problems. BVI constructs a flexible approximating family using a mixture of simple component distributions, with the Gaussian distribution chosen specifically for its ease of optimising and its parametric nature. The components are optimised sequentially using a greedy algorithm, progressively improving the accuracy of the posterior approximation as more components are added. We have demonstrated the effectiveness of BVI through applications to seismic travel time tomography and full waveform inversion (FWI). By comparing the results obtained using BVI with other variational and Monte Carlo sampling methods, we conclude that BVI is capable of providing efficient and accurate inversion results. One key advantage of BVI is its ability to provide an analytic expression for the posterior probability distribution function, which provides a low number of representative samples that partially represent the posterior uncertainty. We have introduced a probabilistic framework that uses these samples to solve an interrogation problem - answering a specific scientific question by interrogating the probabilistic inverse problem solution. The result demonstrates that the representative samples yield similar accuracy compared to that obtained using the full posterior distribution. This approach reduces the computation for subsequent uncertainty analysis, making it promising for large scale problems.

\section*{Acknowledgements}
We thank Edinburgh Imaging Project (EIP) sponsors (BP and TotalEnergies) for supporting this research. AC thanks Muhammad Atif Nawaz for contributory discussions. For the purpose of open access, we have applied a Creative Commons Attribution (CC BY) licence to any Author Accepted Manuscript version arising from this submission.

\bibliographystyle{plainnat}
\bibliography{reference}

\appendix

\section{Derivation and calculation for $\nabla$ELBO}
In this Appendix, we derive the gradient of $\text{ELBO}[q^t(\mathbf{m})]$ with respect to the weight coefficient $w_t$ in equation \ref{eq:weight_2} and the numerical method used for its calculation.

Substitute equation \ref{eq:mix_2pdf} into \ref{eq:elbo}, and this gradient term can be written as
\begin{equation}
	\begin{split}
		\nabla_{w_t} \text{ELBO}[q^t(\mathbf{m})] & = \nabla_{w_t}\mathbb{E}_{q^t(\mathbf{m})}[\log p(\mathbf{m}, \mathbf{d}_{obs}) - \log q^t(\mathbf{m})] \\
		& = \nabla_{q^t}\mathbb{E}_{q^t(\mathbf{m})}[\log p(\mathbf{m}, \mathbf{d}_{obs}) - \log q^t(\mathbf{m})] \nabla_{w_t}\left((1-w_t)q^{t-1}(\mathbf{m})+w_t g_t(\mathbf{m})\right) \\
		& = \int \large(\log p(\mathbf{m}, \mathbf{d}_{obs}) - \log q^t(\mathbf{m})\large) (g_t(\mathbf{m}) - q^{t-1}(\mathbf{m})) d\mathbf{m} \\
		& = \mathbb{E}_{g_t(\mathbf{m})}[\log \dfrac{p(\mathbf{m}, \mathbf{d}_{obs})}{q^t(\mathbf{m})}] - \mathbb{E}_{q^{t-1}(\mathbf{m})}[\log \dfrac{p(\mathbf{m}, \mathbf{d}_{obs})}{q^t(\mathbf{m})}],
	\end{split}
	\label{eq:grad_elbo_w}
\end{equation}
which can be estimated using Monte Carlo integration by drawing samples from $g_t(\mathbf{m})$ and $q^{t-1}(\mathbf{m})$. Then we iteratively update $w_t$ using stochastic gradient descent (equation \ref{eq:weight_2}).

\label{ap:A}

\label{lastpage}
\end{document}